\documentclass[usenatbib]{mnras}

\usepackage{graphicx}
\usepackage{amssymb}
\usepackage{epstopdf}
\usepackage{amsmath}
\usepackage{comment,xcolor}
\usepackage{hyperref}
\usepackage{gensymb}
\usepackage{xspace}
\usepackage{array}
\usepackage{cuted}
\usepackage{orchidlink}
\usepackage{tikz}
\newcommand{\tikzcircle}[2][red,fill=red]{\tikz[baseline=-0.5ex]\draw[#1,radius=#2] (0,0) circle ;}%

\definecolor{dg}{rgb}{0.0, 0.6, 0.1}

\newcommand{\bfm}[1]{\mbox{\boldmath$ #1 $}}

\DeclareMathOperator{\sech}{sech}

\DeclareRobustCommand{\VAN}[3]{#2}
\let\VANthebibliography\thebibliography
\def\thebibliography{\DeclareRobustCommand{\VAN}[3]{##3}\VANthebibliography}

\title{Galactic halo bubble magnetic fields and UHECR deflections}

\author[V.~Shaw et al.]{
Vasundhara~Shaw,$^{1,2}$\thanks{E-mail: vasundhara.shaw@desy.de}\orcidlink{0000-0002-5824-7191},
Arjen~van~Vliet,$^{1,3}$\orcidlink{0000-0003-2827-3361},
Andrew M. Taylor$^{1}$\orcidlink{0000-0001-9473-4758}
\\
$^{1}$Deutsches Elektronen-Synchrotron, Platanenallee 6, Zeuthen, Germany \\ 
$^{2}$University of Potsdam, Institute of Physics and Astronomy, 14476 Potsdam, Germany \\
$^{3}$Department of Physics, Khalifa University, P.O. Box 127788, Abu Dhabi, United Arab Emirates
}
\date{Accepted XXX. Received YYY; in original form ZZZ}

\pubyear{2022}

\begin{document}
\maketitle

\begin{abstract}
We consider the synchrotron emission from electrons out in the Galactic halo bubble region where the Fermi bubble structures reside. Utilising a simple analytical expression for the non-thermal electron distribution and a toy magnetic field model, we simulate polarised synchrotron emission maps at a frequency of 30~GHz. Comparing these maps with observational data, we obtain constraints on the parameters of our toy Galactic halo bubble magnetic field model. Utilising this parameter value range for the toy magnetic field model, we determine the corresponding range of arrival directions and suppression factors of ultra high energy cosmic rays (UHECRs) from potential local source locations.

We find that high levels of flux suppression (down to 2\%) and large deflection angles ($\geq 80^{\circ}$) are possible for source locations whose line-of-sight pass through the Galactic halo bubble region. We conclude that the magnetic field out in the Galactic halo bubble region can strongly dominate the level of deflection UHECRs experience whilst propagating from local sources to Earth.
\end{abstract}


\section{Introduction}
\label{Introducion}

The origin and structure of the Galactic magnetic field remains a long standing unresolved problem in astrophysics. What has become apparent, however, is the vital role it plays, especially in terms of cosmic ray propagation within the Galaxy. The incompleteness of the observational data, required to probe the Galactic magnetic field structure on many different length scales, limits significantly our ability to describe cosmic ray propagation through the Galaxy. This is especially true when it comes to the modelling of cosmic ray propagation out in the Galactic halo region where our knowledge of the magnetic field is particularly weak.

Magnetic fields in the Galactic halo region are primarily probed by two physical processes: Faraday rotation measure in which the thermal electron distribution couples with the line of sight magnetic field ($B_{\parallel}$), and synchrotron radiation in which the non-thermal electron distribution couples to the magnetic field component perpendicular to the line of sight ($B_{\perp}$).

Synchrotron emission from extended Galactic structures at high Galactic latitudes were initially discovered in WMAP data \citep{WMAP_2004}, which were coined the WMAP haze, and later also confirmed by Planck (Planck haze) \citep{Planck_Haze_2013}. This haze emission was revealed to have a bubble-like shape using ground-based S-PASS (radio) observations \citep{Carretti_2013} at 2.3~GHz. A higher energy counterpart of these radio bubbles was in fact earlier discovered by Fermi-LAT at gamma-ray energies (Fermi bubbles), which extend up to $\approx$~3~kpc radially and $\approx$~8~kpc in the z-direction, having a total energy of $\approx 10^{(54-55)}$~ergs \citep{Dobler_2010, Su_2010, Su_2012,Yang_2014,Ackermann_2014}. Spatial correlations between the radio haze and the Fermi bubbles suggest that the origin of these two non-thermal emission components is connected \citep{Su_2010,Crocker_2015}.

Recently, new thermal emission observations made by eROSITA \citep{eROSITA} at X-ray energies have indicated the existence of hot gas in even larger bubble-like structures, extending out to  $\approx$~7~kpc radially and $\approx$~14~kpc in the azimuthal direction. The thermal energy content in these extended bubbles is estimated to be $\approx 10^{56}$~ergs. Collectively, these new non-thermal and thermal observations strongly motivate new investigations into the magnetic fields present in these Fermi/eRosita bubble regions. Henceforth, for the sake of simplicity, we refer to these bubbles together as the Galactic halo bubbles.

Using the argument of equipartition of energy between cosmic rays and magnetic fields \citep{Longair}, one can estimate the strength of the magnetic field in different parts of the Galaxy. Utilising such energy arguments, field strengths in the halo bubbles between $6-10~\mu $G were inferred from the S-PASS observations. The actual value determined here is dependent on the assumed proton-electron ratio in the minimum energy calculation. S-PASS observations, however, are subject to depolarisation of polarised synchrotron radiation via Faraday rotation due to its relatively low observation frequencies. Additionally, this data set is not sensitive to the full portion of the Fermi bubble region of the sky due to the ground-based location of the instrument, restricting observations to only one (the southern) terrestrial hemisphere. For this reason, data from Planck and WMAP offer a more complete probe of the magnetic fields in the Galactic halo bubbles due to their all-sky coverage and observation bandwidths which are not sensitive to Faraday rotation effects. 

Knowledge of the non-thermal electron distribution is a critical ingredient for the determination of synthetic synchrotron maps for a given Galactic magnetic field model.
Direct information on the distribution of non-thermal (cosmic ray) electrons at Earth can be obtained from cosmic ray detectors using, for example, AMS \citep{AMS_2002, AMS_2014}, CALET \citep{Calet_2017} and DAMPE \citep{Dampe_2017}. However, since we do not have direct knowledge of the non-thermal electron distribution throughout the Galaxy, indirect methods to motivate the non-thermal electron distribution in the Galaxy are called upon. Motivations for a number of different models are considered by the community to describe the spatial distribution of relativistic electrons in the Galaxy; for example, either on theoretical grounds using the GALPROP diffusive transport code \citep{Hammurabi, Orlando_2011} or on more phenomenological grounds as done in the WMAP data analysis \citep{WMAP_Page}.

Efforts have been made to model the magnetic fields in the Galaxy, for example by \citet{Sun_2008,Ruiz_Granados_2010,Jaffe_2010,PT11_2011, Jaffe_2011}. 
The current understanding of the magnetic fields out in the Galactic halo is considerably more limited than the magnetic field in the Galactic disc, due to lack of observational data probing this region of the sky at different frequencies \citep{Han_2017}. Even from this limited observational data for the halo, however, evidence has been found to support the torroidal magnetic field models i.e. axisymmetric \citep{WMAP_Page,Sun_2008,Ruiz_Granados_2010,PT11_2011} which are anti-symmetric in their geometry \citep{Han1997antisymmetric,Han1999pulsar,Sun_2008} (ie. the field is oriented in opposite directions on either side of the Galactic plane) in the Galactic halo. Additionally, observations of X-shaped magnetic fields from external galaxies \citep{Krause_2009,Beck_2009} have motivated similar field halo field models for the Milky Way \citep{Katia_2014}. In particular the widely used JF12 model \citep{JF12} provides a two component (toroidal and X-field) description of magnetic fields in the Galactic halo. However, these authors masked out the Fermi bubble regions of the sky in their evaluation of the model agreement with the data. In contrast, the S-PASS observations of the Fermi bubble regions indicated that the magnetic field strength in this region of the sky was considerable. It therefore appears timely to reconsider the modelling of the Galactic halo, utilising new observational results from these bubble regions.

An understanding of the propagation of cosmic rays is vital for resolving their sources. However, this understanding is limited by our current knowledge about the intervening magnetic fields. Extragalactic cosmic rays (ultra high energy cosmic rays (UHECRs) with energies higher than $10^{18}$~eV) are constituted by charged protons or nuclei, and their original directions are, therefore, scrambled by the magnetic fields in the path between the source and Earth. Different models of the Galactic magnetic field give vastly different predictions for the deflection of UHECRs (see e.g.~\citet{Sun_2008, Sun_2010, PT11_2011, JF12, FARRAR_2014}). Recently, significant anisotropies in the UHECR sky have been discovered \citep{TA_2014, Auger_Starburst2018, ICRC_2019, ICRC_2021,Auger_2022}. Due to the deflections in the Galactic magnetic fields, the interpretation of these results in terms of the localisation of the UHECR sources is extremely hard and hence, knowledge of Galactic magnetic fields is extremely important. 

The structure of this paper is the following. In section~\ref{Methods} we provide a description of the electron distribution and the toy magnetic field model adopted in our study. In section~\ref{Results} synthetic polarised synchrotron maps are produced adopting this model, which are then compared against the Planck data. A grid scan of the model against the data is then made in order to obtain constrained model parameters. In section~\ref{Deflections} we determine the arrival directions of ultra high energy cosmic rays with $E = 40$~EeV from our toy model and discuss how the uncertainties in the parameters can propagate errors in estimating the cosmic ray deflections. 
Lastly, in section~\ref{Conclusions} we summarise our conclusions.

\section{Galactic Halo bubble Magnetic Field Model}
\label{Methods}

\subsection{Toy Model for the Galactic Halo Bubbles}
\label{GMF}
In this paper we follow the philosophy of \citet{West_Helicity}, adopting a simple toy model as means of a preliminary attempt to provide a model for the Galactic halo bubbles. 

\begin{figure*}
\centering
\includegraphics[width = 0.49\linewidth]{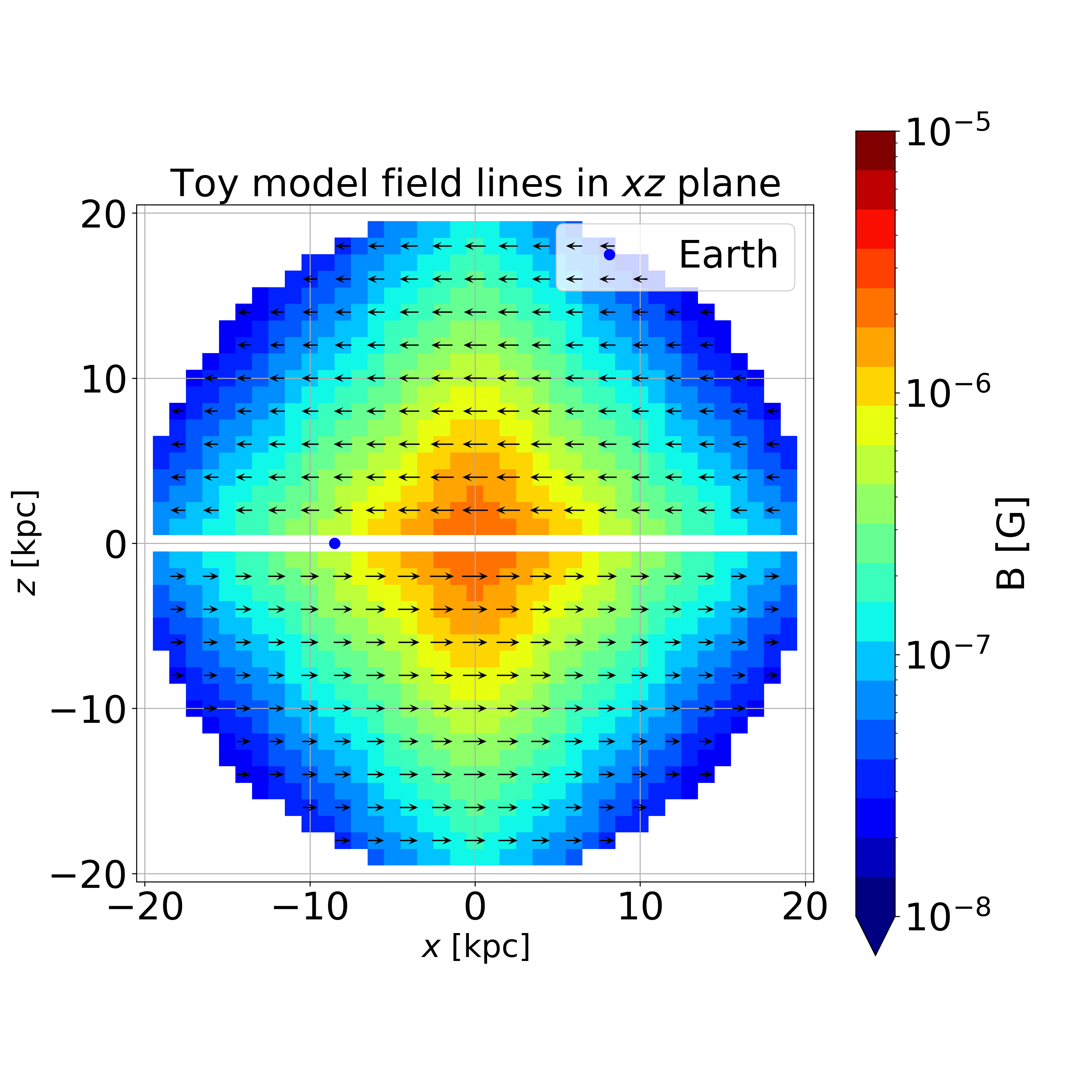}%
\includegraphics[width = 0.49\linewidth]{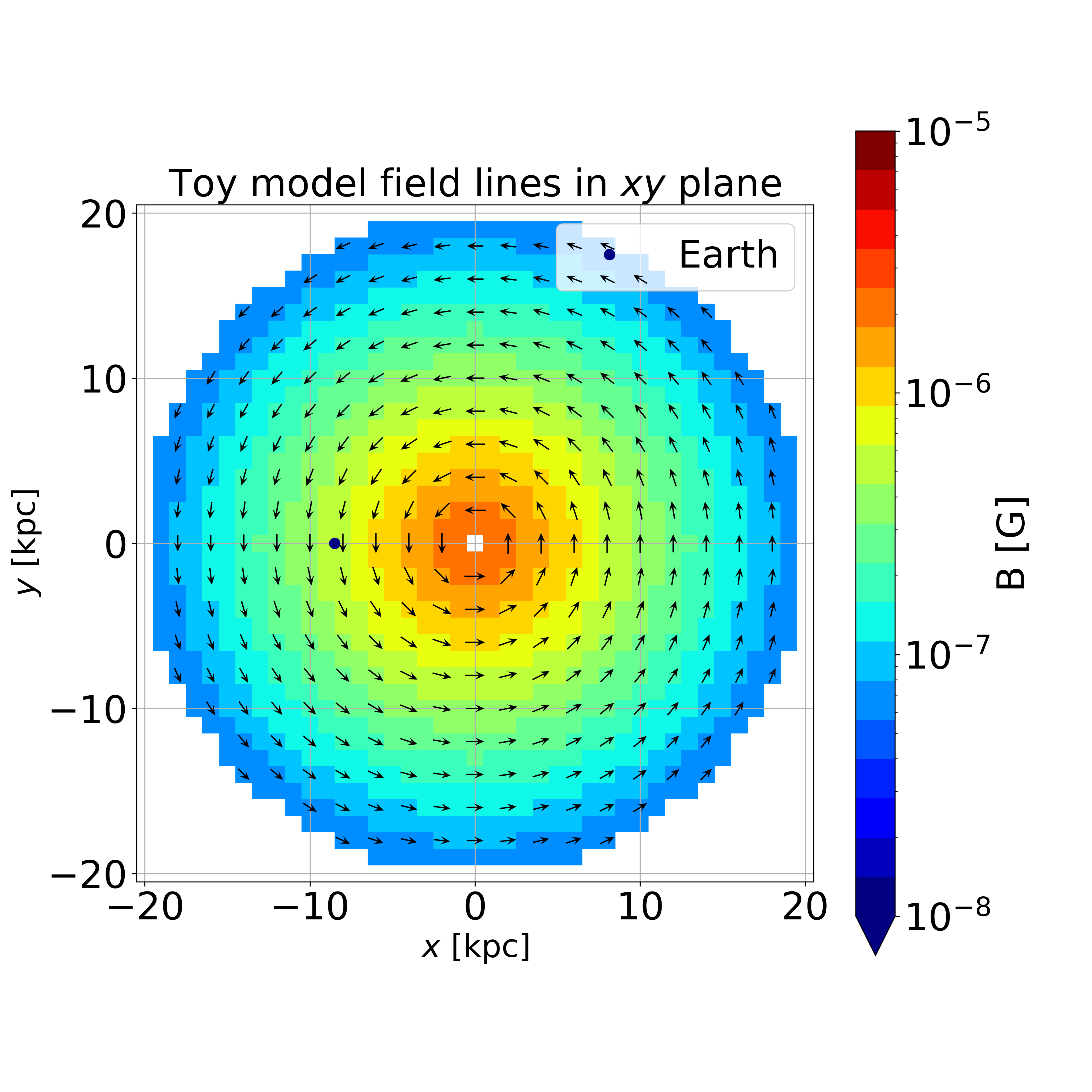}
\caption{Cross-section of the toy model for the Galactic magnetic field (for best fit parameter values see Table~\ref{Para_table}) in the Galactic halo bubble region in the $xy$ plane at $z = 1$~kpc and $xz$ plane at $y = 1$~kpc (with the Galactic plane in the $xy$ plane at $z=0$) showing their drop in two dimensions. We omit the disc region in the \textbf{left} plot since its not a part of our model.}
\label{fig:Vis_TM}
\end{figure*}

For our toy model, we adopt an axisymmetric toroidal structured field along with an additional turbulent field component. The strength of the toroidal structured field is described by:
\begin{equation}\label{TM_equation}
B_{\rm{tor}}(r,z) = B_{\rm{str}} {\rm{e}}^{-|z|/Z_{\rm{mag}}} {\rm{e}}^{-z_{\rm{min}}/|z|} {\rm{e}}^{-r/R_{\rm{mag}}},
\end{equation}
with $r$ the radial distance from the Galactic centre in the $xy$ plane.
The structured field has 3 free parameters: $B_{\rm str}$ as the strength of the magnetic field and $R_{\rm {mag}}$ and $Z_{\rm {mag}}$ describing the radial and azimuthal cut off distances, respectively. The value of $z_{\rm min}$ = 100~pc, which dictates the cut in the Galactic plane, is fixed. The model calculations are continued up to 14~kpc from the Galactic centre with the observer being centered at Earth, (-8.5,0,0)~kpc. The direction of the toroidal field is orientated in opposite directions above and below the Galactic plane. A visualisation of our magnetic field in \textit{xy} and \textit{xz} cross-sections is shown in Fig.~\ref{fig:Vis_TM}. 

For the turbulent fields we use a 5/3 Kolmogorov power-law spectrum with a mean RMS value for this component of $B_{\rm{tur}}$.
We use CRPropa~3 \citep{CRPropa3_2016} for generating these turbulent fields. 
The minimum and maximum values of the wavelength to generate these fields are  $L_{\rm min}$ = 200~pc and $L_{\rm max}$ = 400~pc. For computational reasons we stick to this restricted dynamic range of $L_{\rm min}$ and $L_{\rm max}$. In Appendix~\ref{Appendix_A} we discuss the effect of this small dynamic range in detail. The turbulent field has effectively only 1 free parameter which is the magnitude of the turbulent field strength, $ B_{\rm tur}$, with the coherence length of the field being kept fixed at 150~pc. This value of $L_{\rm coh}$, although very large, does sit in the range of values considered \citep{Ohno_1993, Chepurnov_2010, Haverkorn_2013,Beck_2016, Giacinti_2018}. The turbulent fields in our model extend only out to 14~kpc radially from the Galactic centre, chosen so as to encompass the Galactic bubble region reported \citep{eROSITA}.
In Appendix~\ref{Appendix_B} we show a power-spectrum plot for the actual turbulent magnetic field realisation adopted. 

Since we focus only on the regions of the sky which probe the Galactic halo bubbles we do not include any disc magnetic field component in this model. For the purposes of comparison, we use the JF12 model as a reference since it is a widely known Galactic magnetic field model.
However, it should be noted that the JF12 model was motivated by observations which masked out a large part of the Galactic bubble region of the sky that we focus on, and adopts magnetic field strengths and spatial extensions both weaker and smaller than those suggested by the S-PASS observations \citep{Carretti_2013} of these bubble regions.

\subsection{Non-Thermal Electron Distribution}

\begin{figure*}
\centering
\includegraphics[width=0.49\linewidth]{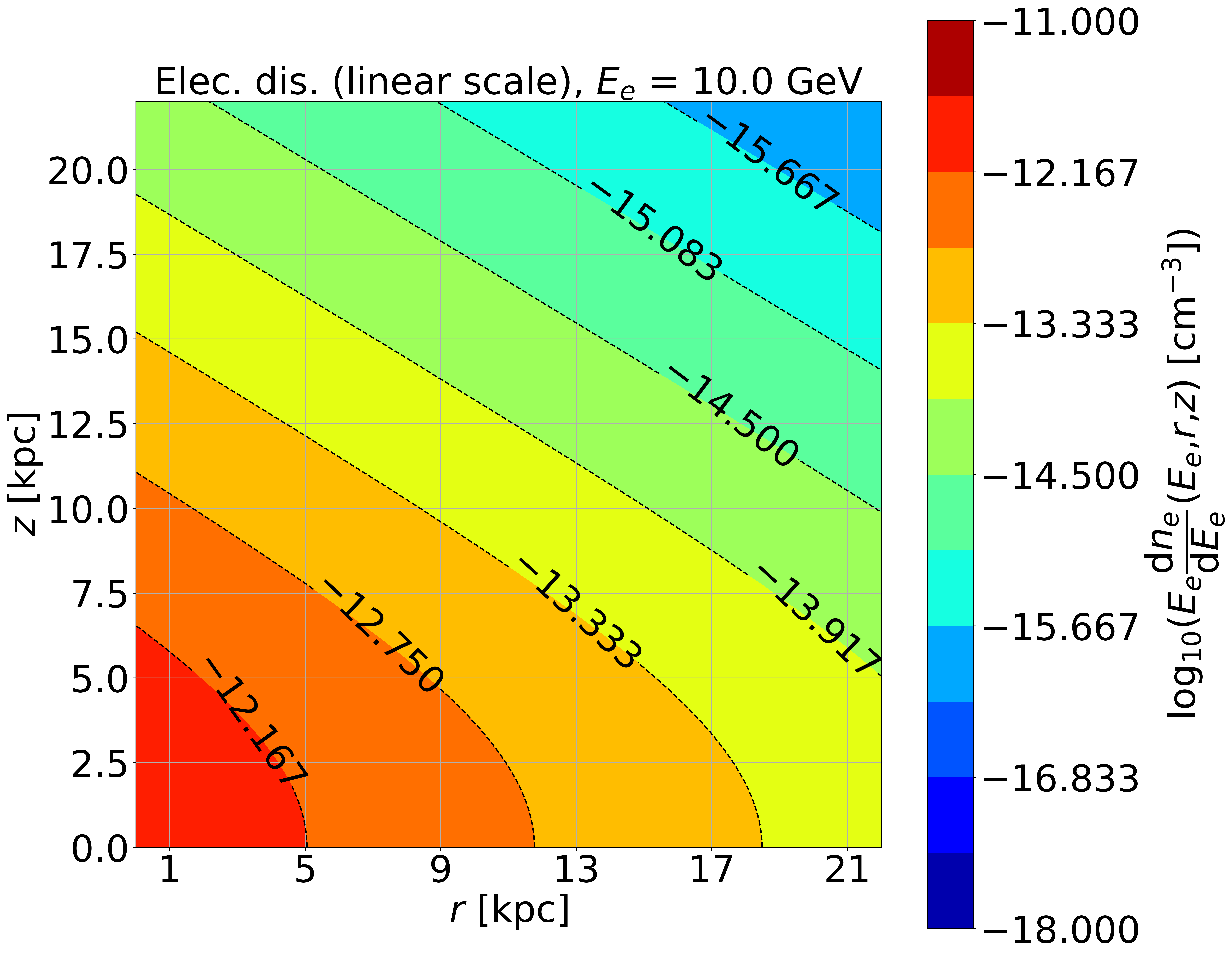}%
\includegraphics[width = 0.49\linewidth]{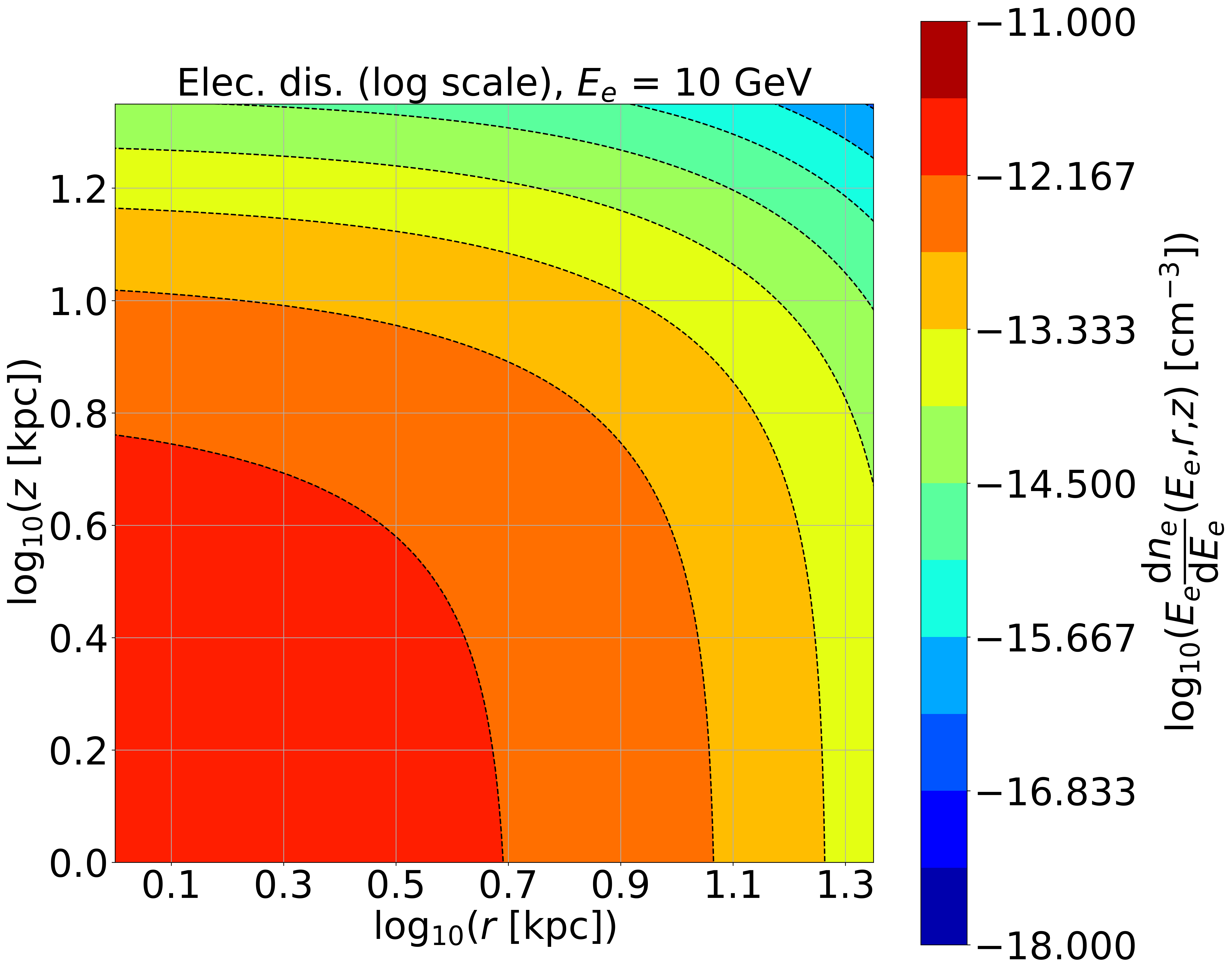}
\caption{An example of the electron distribution for $E_e = 10$~GeV, $R_{\mathrm{el}} = 5$~kpc and $Z_{\mathrm{el}} = 6$~kpc in linear scale on the left and in log-scale on the right. The $C_\mathrm{norm}$ value for this plot is $10^{-11.72}~{\rm cm}^{-3}$ (see Table~\ref{Para_table}).}  
\label{fig:electron_density}
\end{figure*}

In order to calculate synthetic synchrotron maps, both a non-thermal electron distribution and magnetic field model are required. For the non-thermal electron distribution, the JF12 model considered both the WMAP analytical expression (Eq.~\ref{Eq_WMAP_EdNdE}) and a simulated non-thermal electron distribution from GALPROP, with the latter being adopted. These two non-thermal electron models are quite different. The WMAP model \citep{WMAP_Page} is an analytical expression whereas the GALPROP distribution \citep{Hammurabi} is more theoretical in motivation, being obtained from a solution to the diffusive transport equation assuming a specific spatial distribution for the sources, with an absorptive halo boundary. As our current knowledge of the non-thermal electron distribution in the Galaxy, especially in the Galactic halo bubble region, is very limited, we choose to adopt the simple WMAP analytical model in order to avoid adding further layers of complexity. The WMAP non-thermal electron density distribution model we adopt has the form:
\begin{equation}\label{Eq_WMAP_EdNdE}
    \frac{\mathrm{d}n_e}{\mathrm{dlog}E_{e}} =     C_\mathrm{norm} \left(\frac{E_e}{E_{\rm 10GeV}}\right)^{-p+1} e^{-r/R_{\mathrm{el}}} \sech^2\left(\frac{z}{Z_{\mathrm{el}}}\right), 
\end{equation}
where $\frac{\mathrm{d}n_e}{\mathrm{dlog}E_{e}}$ is the differential electron spectrum in logarithmic energy bins, in units of ${\rm cm}^{-3}$, and $p =3$ is the spectral index of the electron spectrum. The parameter $C_\mathrm{norm}$ describes the electron density for electrons with an energy of 10~GeV, and $R_{\mathrm{el}}$ \& $Z_{\mathrm{el}}$ describe the radial and azimuthal spatial cut-offs. For reference, in Fig.~\ref{fig:electron_density} we show a spatial distribution of 10~GeV electrons both in linear and logarithmic space.

It should be noted that in our description of the halo, it is assumed that both the magnetic field and electron distribution possess an exponential cut-off in their spatial extent beyond a cut-off distance scale, whereas in reality they may have a power-law decay beyond this distance
\citep{Hammurabi, Subramanian_2018, Bell_2022}. However, since we are primarily interested in regions dominating the total synchrotron emission, the actual distribution of the particles and field beyond the scale height distance are not our focus. Provided that the synchrotron emissivity decays faster than $l^{-1}$ along the line of sight at distances beyond the cut off distance, the contribution to the synchrotron emission from further distances can be safely neglected. 

\subsection{Synchrotron Emission}\label{Synchrotron_theory}

\subsubsection{Intensity \& polarisation}
Synchrotron radiation or magneto-bremsstrahlung radiation is the radiation produced due to charged particles that gyrate at relativistic speeds around a static magnetic field. Synchrotron radiation is sensitive to $B_{\perp}$, the magnetic field component perpendicular to the line of sight. The radiation produced via synchrotron is often linearly polarised.
The polarised emissivity (emission per unit volume) spectral distribution can be visualised as an ellipse where the major axis is the perpendicular component ($J_{\rm \perp}$) and the minor axis is the parallel ($J_{\parallel}$) component (see Appendix~\ref{Appendix_C} for further discussion). 
The two polarisation emission components, $J_{\perp}$ and $J_{\parallel}$, describe the emission spectrum for a given peak photon energy $E_{\gamma}^{\mathrm{peak}}$.
Expressions for these two components, produced by electrons with pitch angle $B_{\perp}/B$, are provided below in Eqs.~\ref{Jperp} and \ref{Jpara},

\begin{equation}
 {J_{\perp}^l} = \frac{1}{\tau}  \int_{\mathrm{log}E_e^{\mathrm{min}}}^{\mathrm{log}E_e^{\mathrm{max}}} \  \frac{\mathrm{d}n_e}{\mathrm{dlog}E_{e}} \mathrm{dlog}E_{e}\  \left[F\left(\frac{E_{\gamma}}{E_{\gamma}^{\mathrm{peak}}}\right) + G\left(\frac{E_{\gamma}}{E_{\gamma}^{\mathrm{peak}}}\right)\right] \
 \label{Jperp}
\end{equation}

\noindent and

\begin{equation}
{J_{\parallel}^l} = \frac{1}{\tau} \int_{\mathrm{log}E_e^{\mathrm{min}}}^{\mathrm{log}E_e^{\mathrm{max}}} \ \frac{\mathrm{d}n_e}{\mathrm{dlog}E_{e}} \mathrm{dlog}E_{e}\  \left[F\left(\frac{E_{\gamma}}{E_{\gamma}^{\mathrm{peak}}}\right) - G\left(\frac{E_{\gamma}}{E_{\gamma}^{\mathrm{peak}}}\right)\right] 
\label{Jpara}
\end{equation}

\noindent where
\begin{align}
\tau^{-1} &= \frac{\sqrt{3} \alpha}{4\pi}\frac{B_{\perp}}{B_{\mathrm{crit}}}\frac{m_{e}c^{2}}{\hbar},
 & 
E_{\gamma}^{\mathrm{peak}} &= \frac{3}{2}\Gamma_{e}^2 \frac{B_{\perp}}{B_{\mathrm{crit}}} m_{e} c^2,
\nonumber
\end{align}
and
\begin{align}
F(x) &= x \int_x^\infty K_{5/3}(x') dx', &
G(x) &= x K_{2/3}.
\nonumber
\end{align}

These expressions are provided in terms of the critical magnetic field strength, $B_{\mathrm{crit}} = \frac{m_e^2c^3}{e\hbar} = 4.414 \times 10^{13}$~G, where $m_e c^{2} = 0.511$~MeV is the rest-mass energy of the electron, $h = 4.136 \times 10^{-15}$~eV~s is Planck's constant, $\Gamma_{e}$ is the electron Lorentz factor and $\alpha \approx \frac{1}{137.04}$ is the electromagnetic fine structure constant. 
In the case of a mono-energetic electron energy distribution with density $n_{e}$, we can calculate the total radiated power density by summing Eqs. \ref{Jperp} and \ref{Jpara} and integrating over the photon energy distribution:
\begin{equation}
\frac{{\rm d}E_e^{\rm tot}}{{\rm d}t} = \frac{2\alpha}{3} n_{e}\left(\frac{B_{\perp}}{B_{\rm crit}}\right)^{2}\frac{E_{e}^{2}}{\hbar}
\label{Itot_l}
\end{equation}
where the result $\int_0^{\infty} F(x) \rm{d}x = 8\pi/(9\sqrt{3})$ \citep{Westfold} has been used. The above expressions can be used to compute the pitch angle averaged synchrotron cooling time  [$\langle B_{\perp}^{2}\rangle = (2/3) B^{2}$] for electrons in this unit system, given by $\tau_c = \frac{E_e}{{\rm d}E_e/{\rm d}t}$ \citep{Taylor_Matthews}.

For clarity, several of the conventions we adopted are noted here. The parallel component of polarisation (${J_{\parallel}}$) is orientated in the same direction as  $\vec{B}_{\perp}$, and the perpendicular component of polarisation (${J_{\perp}}$) is perpendicular to $\vec{B}_{\perp}$. The Stokes parameters at each point along the line of sight can be written in terms of the intrinsic polarisation angle $\Psi^l_{\rm in}$, which is the angle between the line-of-sight perpendicular component of the magnetic field $B_{\perp}$ and Galactic south at each step. The conventions adopted here match those used by the Planck collaboration \citep{Planck_XIX} based on the $\rm HEALPix$\footnote{\textcolor{purple}{https://healpix.jpl.nasa.gov/}} software by \citet{Healpix_2005}. For each step along the line of sight, both  ${J_{\perp}^l}$ and ${J_{\parallel}^l}$ are subsequently used to obtain the $Q$ and $U$ Stokes parameters. We obtain values for the intrinsic Stokes parameters $Q^{\rm tot}_{\rm in}$ and $U^{\rm tot}_{\rm in}$ by integrating over $Q$ and $U$ along the line of sight:

\begin{eqnarray}
Q_{\rm in}^{\rm tot} = \frac{1}{4\pi} {\int_0^L \mathrm{d}l \ ({J_{\perp}^l} - J_{\parallel}^l) \ {\cos}(2\Psi^l_{\rm in}) }, \\
U_{\rm in}^{\rm tot} =\frac{1}{4\pi} {\int_0^L \mathrm{d}l \ ({J_{\perp}^l} - J_{\parallel}^l) \ {\sin}(2\Psi^l_{\rm in})}.
\end{eqnarray}

The polarised flux ($I_{\rm pol}$) can then be expressed in terms of ${Q_{\rm in}^{\rm tot}}$ and ${U_{\rm in}^{\rm tot}}$ as
\begin{eqnarray} \label{eq_I_pol}
I_{\rm pol} = \sqrt{(Q_{\rm in}^{\rm tot})^2+(U_{\rm in}^{\rm tot})^2} = J_{\perp}^{\rm tot} - J_{\parallel}^{\rm tot}.
\end{eqnarray}
Similarly, $I_{\rm tot}$ is computed by summing the contributions of $J_{\perp}^l$ and and $J_{\parallel}^l$ for each point along the line of sight,
\begin{equation} \label{eq_I_tot}
    I_{\rm tot} = \frac{1}{4\pi} \int_0^L \mathrm{d}l (J_{\perp}^l + J_{\parallel}^l).
\end{equation}

$J_{\perp}^{\rm tot}$ and $J_{\parallel}^{\rm tot}$ are the resultant magnitudes of emissions in perpendicular and parallel directions and can be given by:
\begin{eqnarray}
J_{\perp}^{\rm tot} = (I_{\rm tot} + I_{\rm pol})/2, \\
J_{\parallel}^{\rm tot} = (I_{\rm tot} - I_{\rm pol})/2. 
\end{eqnarray}
The intrinsic polarisation angle $\Psi_{\rm in}$ is the resulting angle of polarisation:
\begin{eqnarray}
\tan(2\Psi_{\rm in}) = \frac{U_{\rm in}^{\rm tot}}{Q_{\rm in}^{\rm tot}}. 
\end{eqnarray}
In Appendix~\ref{Appendix_C} an example case for these calculations is provided for further understanding.

\subsubsection{Simulation setup for the polarised synchrotron emission}
Utilising the setup described in Section~\ref{Methods}, we generate a synthetic polarised synchrotron emission map for each parameter set of our toy model. The toy model comprises of 5 free parameters, (see Table~\ref{Para_table}). The radial cut off of the magnetic field and electron distribution is kept identical ($R_{\mathrm{Mag}}$ = $R_{\mathrm{el}}$) and the same applies to the azimuthal cut-off ($Z_{\mathrm{Mag}}$ = $Z_{\mathrm{el}}$). The reason for this constraint is that the synchrotron radiation level depends on both the non-thermal electron density and the magnetic field strength. Thus, even if the spatial extend of the magnetic field differs from the electron distribution, one can only probe the magnetic field in the region where both the magnetic field and non-thermal electrons are present. 
For the spatial parameter scan, the parameter values scanned over for $R_{\mathrm{el}}$ and $Z_{\mathrm{el}}$ are 2~kpc to 19~kpc, with a scanning step size of 1~kpc. 
However, the range over which both $B_{\rm str}$ and $B_{\rm tur}$ are scanned is binned logarithmically with 30 bins per decade between 2$~\mu$G and 18$~\mu$G. Likewise, for the $C_{\rm norm}$ we scanned between $10^{-14} \rm cm^{-3}$ and $10^{-11} \rm cm^{-3}$, adopting 10 bins per decade. 


\begin{figure*}
\centering
\includegraphics[width=0.49\linewidth]{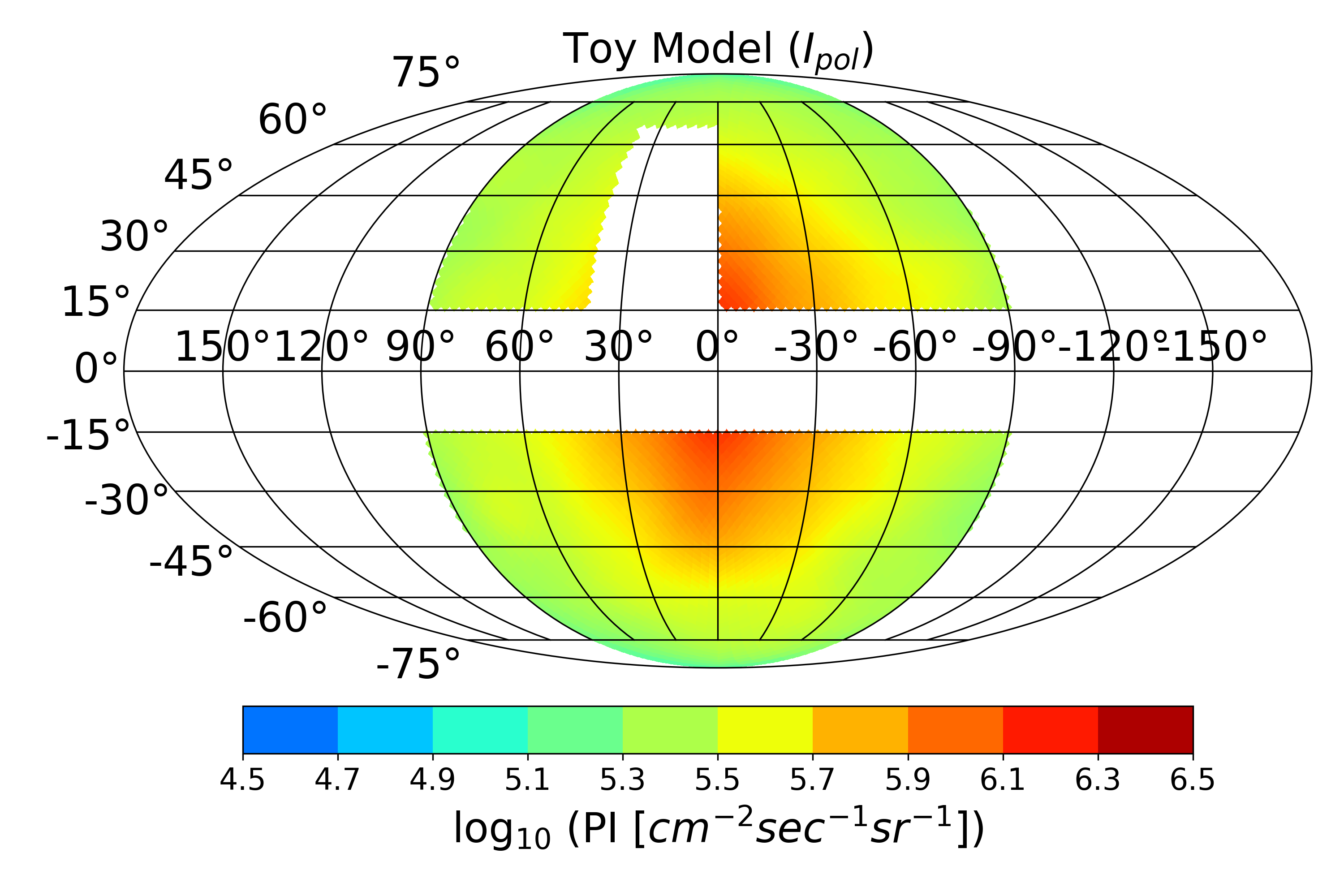}
\includegraphics[width =0.49\linewidth]{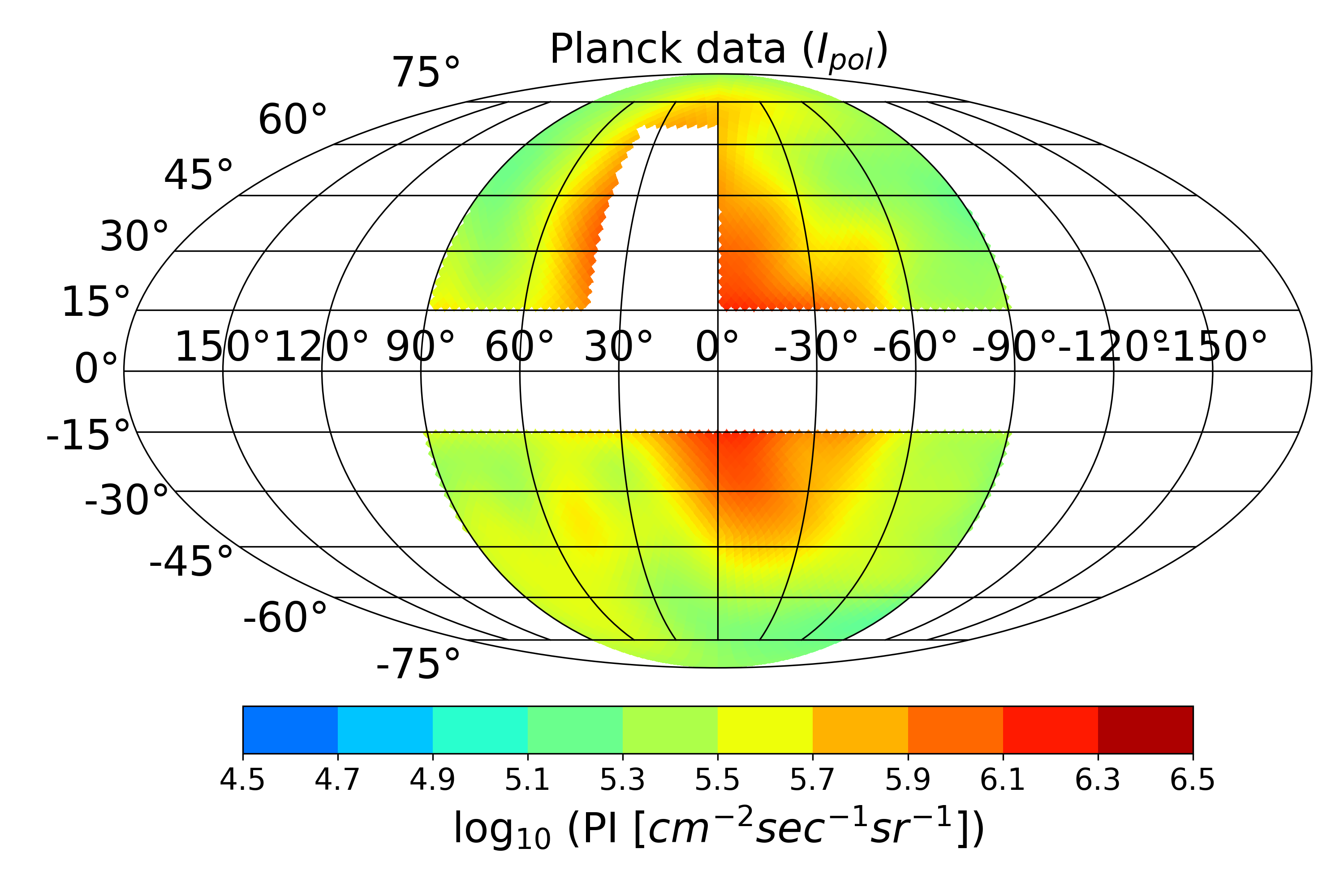}%

\includegraphics[width = 0.49\linewidth]{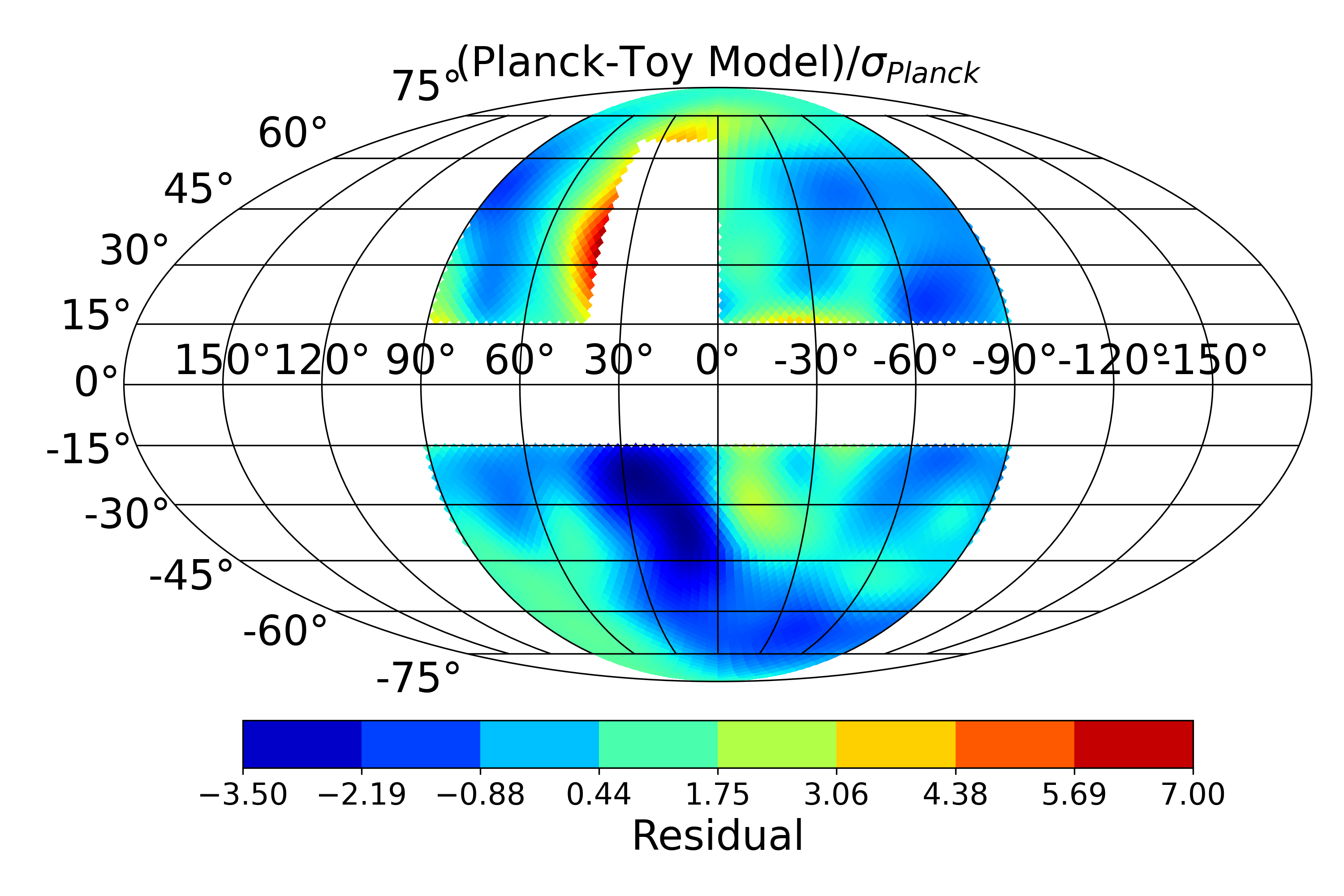}%
\includegraphics[width =0.49\linewidth]{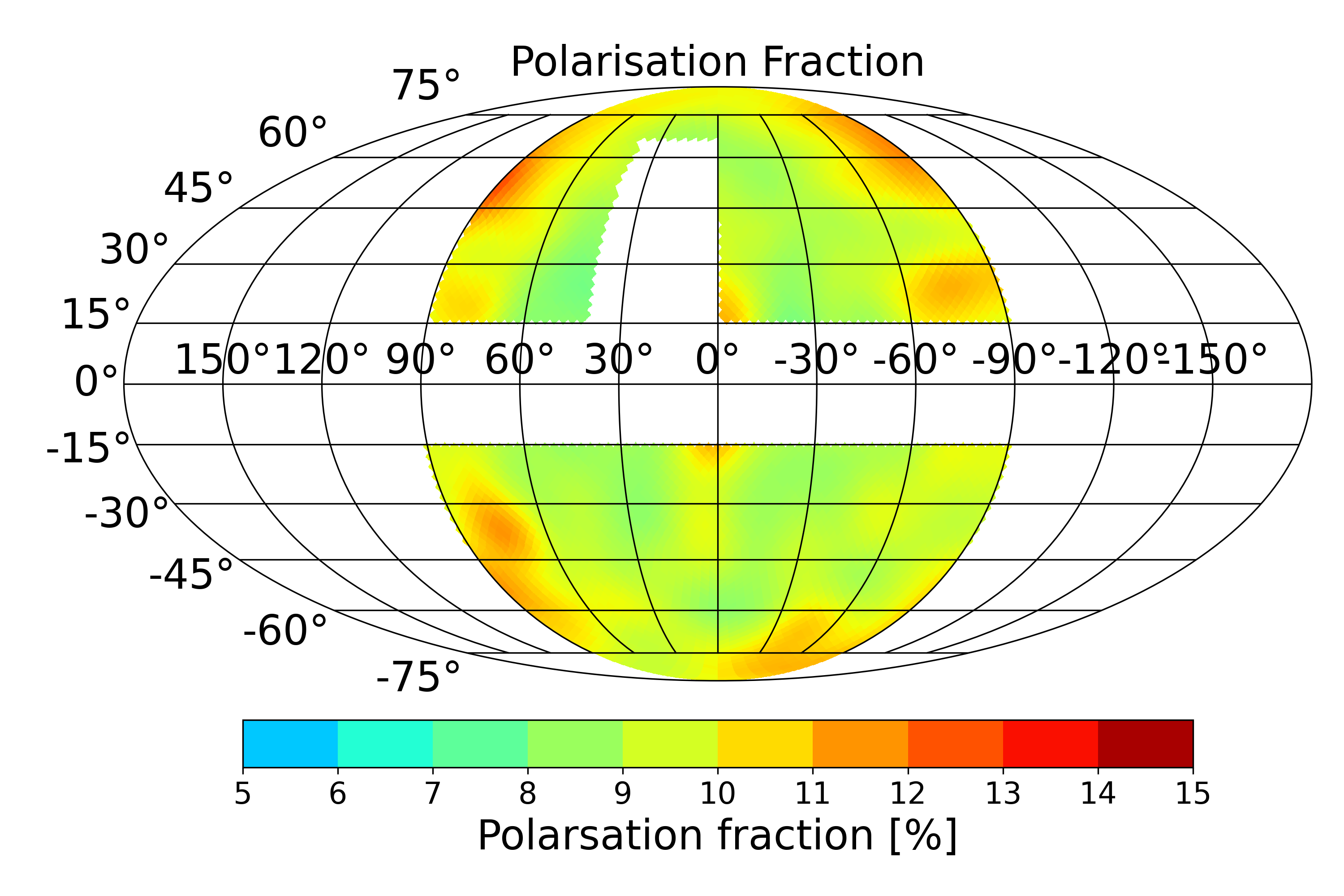}
\caption{\textbf{Top:} Simulated polarised intensity (\textbf{left}) and  Planck polarised intensity skymap (\textbf{right}) for the best-fit parameters (see Table~\ref{Para_table}). \textbf{Bottom:} Residual of the observation and the simulated data (\textbf{left}) and the polarisation fraction for the toy model (\textbf{right}).}
\label{fig:Skymaps}
\end{figure*}

In our study, we mask out three regions of the sky from our skymaps. The first of these is in the Galactic disc region between b = $(-15^{\circ},15^{\circ})$. For the second region, based on observations from \citet{Su_2010} and \citet{eROSITA}, we block out longitudes  $\geq \pm 90^{\circ}$ from the Galactic centre (i.e.~all directions pointing away from the Galactic centre direction), so as to ensure that our analysis only covers the region occupied by the Galactic Halo (Fermi and eRosita) bubbles. Lastly, we block out the region associated with the North Polar Spur (NPS). Our motivation here is that there are indications that the higher latitudes of the NPS are originating locally rather than from the Galactic centre, based on starlight polarisation observations \citep{Gina_2021}. In order to remain as impartial as possible for the designation of this region, we adopt a cut for it selected by \citet{Wolleben_2007}. In Fig.~\ref{fig:Skymaps} observational and synthetic skymaps are shown with these three regions removed.


To obtain the best-fit parameters for our model and their constraints, we ran a grid search over the 5 free parameters, sampling in total $8\times 10^{6}$ parameter sets. For each model parameter configuration, a synthetic skymap was generated using Healpix~\citep{Healpix_2005}, adopting a resolution with Nside = 32. Since the interests of our study are focused on large scale structures, both the synthetic skymaps and observational data were smoothed out, using a Gaussian kernel, on a size scale of $15^{\circ}$, to wash out smaller scale features. We then compare the simulated polarised emission with the Planck data at 30~GHz by evaluating the $\chi^{2}$ value of the model fit to the data.
For this work we consciously decided to carry out the smoothening after calculating polarised emission from the Stokes Q and U maps for both synthetic data and observational data. Our future plan is to improve this method and look into other ways to compare observational and synthetic data.

\begin{table*}
\centering
\caption{Table of best-fit parameters with uncertainties}
\begin{tabular}{ |p{4.cm}|p{4.5cm}|p{6.5cm}|  }
\hline
\multicolumn{3}{|c|}{Best-fit value with 1$\sigma$ constraint} \\
\hline
\rule{0pt}{3ex}
Parameter & Best-fit value &Description \\
\hline
\hline
\rule{0pt}{3ex}
$B_{\mathrm{str}} $& $3.96_{-1.96}^{+6.63} ~ \mu$G & Structured field strength \\
\hline
\rule{0pt}{3ex}
$B_{\mathrm{tur}} $& $ 6.72_{-3.56}^{+9.97} ~\mu$G & Turbulent field strength\\
\hline
\rule{0pt}{3ex}
$R_{\mathrm{Mag}}$ = $R_{\mathrm{el}}$ & $5_{-0}^{+1}$~kpc & Radial cut off \\
\hline
\rule{0pt}{3ex}
$Z_{\mathrm{Mag}}$ = $Z_{\mathrm{el}}$ & $6_{-0}^{+1}$~kpc & Azimuthal cut off\\
\hline
\rule{0pt}{3ex} 
${\rm{log_{10}}}~(C_{\rm norm} [{\rm cm}^{-3}]$) & ${-11.72}_{{-0.93}}^{{+0.62}}$ & Electron {density} at 10~GeV\\
\hline
\end{tabular}
\label{Para_table}
\end{table*}

\subsubsection{Observational data}
For our synchrotron emission study, we use the publicly available data from the Planck satellite mission\footnote{\textcolor{purple}{http://pla.esac.esa.int/pla/}}. Specifically, we use the polarised radio data at 30~GHz from Planck where the peak frequency is at 28.4~GHz, with a band width of 9.8~GHz. At this frequency a considerable level of polarised synchrotron emission is observed, with only a small level of Faraday rotation occurring at these high frequencies. However, we also note that in this 30~GHz band, the Planck data cannot be used to probe synchrotron intensity directly, since at this frequency the unpolarised sky receives considerable contributions from both thermal bremsstrahlung and anomalous microwave emission, as well as synchrotron radiation \citep{Planck_XIX, Planck_X, Planck_XXV, Planck_XLII}. 

\subsection{Constraints on the Magnetic Field Model}
\label{Results}
We obtain 1$\sigma$ constraints on each of our model parameters (see Table~$\ref{Para_table}$). For the structured magnetic field strength, $B_{\rm str}$, we obtain the best-fit value of 3.96~$\mu$G with the upper extreme being 10.59~$\mu$G and the lower extreme 2~$\mu$G. Similarly, for the turbulent magnetic fields, $B_{\rm tur}$, the mean value is 6.72~$\mu$G with lower and upper extreme values of 3.15~$\mu$G and 16.69~$\mu$G, respectively. For the spatial extent of the field, we obtain a best-fit vale of  5~kpc and 6~kpc for the radial ($R_{\rm Mag}/R_{\rm el}$) and azimuthal extent ($Z_{\rm Mag}/Z_{\rm el}$) respectively. We find an upper extreme value of $+ 1$~kpc for the spatial extent whereas the lower extreme remains the same as best-fit value. This is likely an effect of having a large step size in the parameter scan. In case of the non-thermal (10~GeV) electron {density} ${\rm log_{10}} (C_{\rm norm} [{\rm cm}^{-3}])$, the best fit value obtained is -11.72 with upper and lower extreme values being -11.0 and -12.65, respectively. The constraint values for all parameters are in agreement with the observations made by Fermi~\citep{Su_2010}, S-PASS~\citep{Carretti_2013} and eROSITA~\citep{eROSITA}. The dominance of turbulent to structured fields are consistent with the findings from studies of other local galaxies \citep{Beck_NGC_6946,Tabatabaei_2008}.

In Fig.~\ref{fig:Skymaps} the smoothened skymap obtained from the best-fit values of the parameters and the smoothened polarised Planck data are shown along with the residuals. The best-fit values used for the parameters are provided in Table~\ref{Para_table}. 
The polarisation fraction obtained by our best-fit toy-model, given in Fig.~\ref{fig:Skymaps}, was calculated taking the ratio of the polarised to the total intensity. The polarisation fraction for the best-fit toy model is comparable to the values as seen in the observation data of \citet{WMAP_Page} and \citet{Carretti_2013}.

We obtain a ${\chi^2}/{\rm d.o.f.}=$~1.7 from our toy-model when compared with Planck 30~GHz polarised emission data. For the same analysis, the JF12 full halo (no disc) model yields
a  ${\chi^2}/{\rm d.o.f.} = $~6.0 and the Xu \& Han halo model (XH19) \citep{Han_2019} gives a ${\chi^2}/{\rm d.o.f.} = $~11.0  (see Appendix \ref{Appendix_F}). The low latitude polarised synchrotron emission from the JF12 full halo and the XH19 model drives their poor ${\chi^2}/{\rm d.o.f.}$ value (see Fig. \ref{fig:Skymaps_XH19_JF12}). 
Such low latitude emission occurs either due to a very weak turbulent magnetic field model being adopted, as seen in the case of the JF12 full halo model or the complete lack of turbulent fields as seen in the XH19 model. The lack of polarised synchrotron emission at high latitudes obtained from the JF12 model has also been addressed previously by \citet{Beck_2016}.
In comparison, our toy model for the Galactic halo bubble region is statistically significantly better at describing the polarised emission seen by Planck 30~GHz map for the sky region focused on (see Fig. \ref{fig:Skymaps}).

\section{Cosmic ray deflections due to the magnetic field model}
\label{Deflections}

Charged particles propagating through magnetic fields precess around the field lines by virtue of the Lorentz force 
\begin{eqnarray}
\frac{{\rm d}\bfm{\beta}}{c{\rm d}t} = \frac{1}{r_{L}}\bfm{\beta}\times \bfm{\hat{B}}, 
\end{eqnarray}
where $\bfm{\beta}$ is the particle's velocity vector, $\bfm{\hat{B}}$ is the magnetic field unit direction vector, and $r_{L}$ is the particle's Larmor radius. The particle's Larmor radius is defined by $r_{L}=pc/ZeB=R/B$, where $R=E/eZ $ is the particle's rigidity and $Z$ is the nucleus's proton number. 

UHECRs experience deflection effects when propagating through both extragalactic and Galactic magnetic fields. The extragalactic magnetic field is considered to be weak, with $B < \rm {nG}$ for $\lambda_{\rm coh}=1$~Mpc \citep{Blasi_1999, Kronberg_2007}. For UHECRs with rigidity $R > 10^{19}$~V in weak (sub nG) extragalactic magnetic fields, $r_{L}>{\rm 10~Mpc}$, giving rise to a deflection of $\theta\approx \lambda_{\rm coh}/r_{L}<6^{\circ}$ each coherence length. Thus, the angular deflection expected from UHECR propagating from local ($<4$~Mpc) sources a few coherence lengths away is $\lesssim 10^{\circ}$. In comparison, within the Galactic magnetic field structure, field strengths of order $5~\mu$G are experienced. An UHECR with rigidity 10~EV in a $5~\mu$G field, has a Larmor radius of $r_L \approx 2 ~ \rm kpc$. Thus, UHECRs in this rigidity range from a nearby source will pick up their largest angular deflections from their source positions upon passing through the large-scale Galactic magnetic field region.

We use the publicly available cosmic ray propagation code CRPropa~3~\citep{CRPropa3_2016} for studying the effects of toy model magnetic fields on the arrival directions of cosmic rays. Within this software we use the Boris pusher scheme in order to ensure a particle's trajectory evolution satisfies the Lorentz force equation. It is important to note that CRPropa conserves the total energy of each particle during the propagation.

We propagate $4 \times 10^7$ nitrogen cosmic rays with rigidity $R = 6$~EV starting at Earth isotropically through the toy model using the backtracking scheme out to the edge of the simulation box, to a distance of 30~kpc from the Galactic centre. This dimension of the simulation box was chosen such that the results remain insensitive to it, with the Larmor radius of the particle at this distance being an order of magnitude larger than the box size, for the strongest magnetic field case considered. The choice of species for nitrogen is based on composition measures from the Pierre Auger Observatory (PAO) \citep{Auger_2014} and the choice of rigidity was based on arrival-direction correlations with nearby galaxies detected by the PAO \citep{Auger_Starburst2018,Auger_2022}.

\subsection{Effects of the Magnetic Field Model on UHECR Arrival Directions}
\begin{figure*}
\centering

\includegraphics[width=0.49\linewidth]{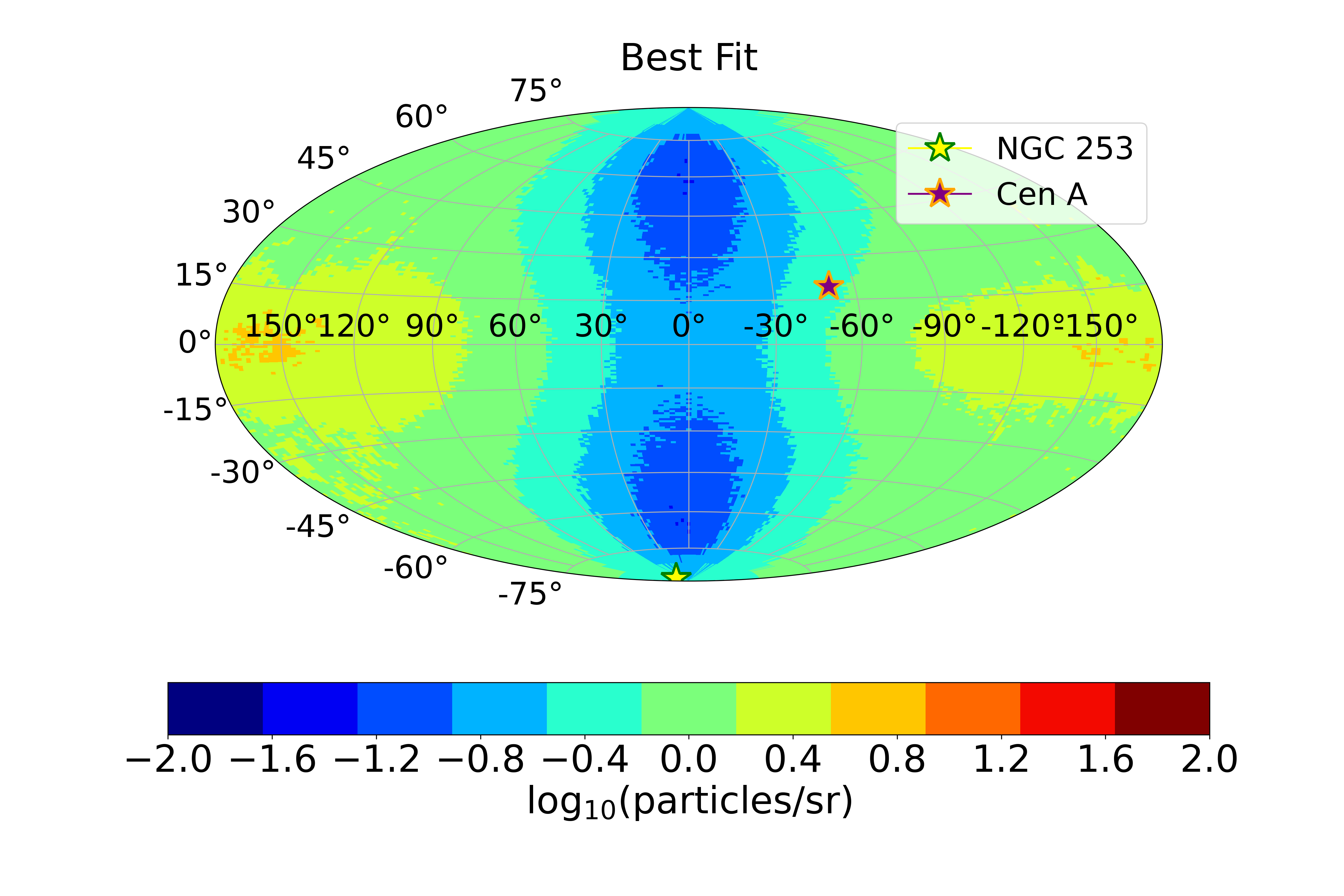}
\includegraphics[width=0.49\linewidth]{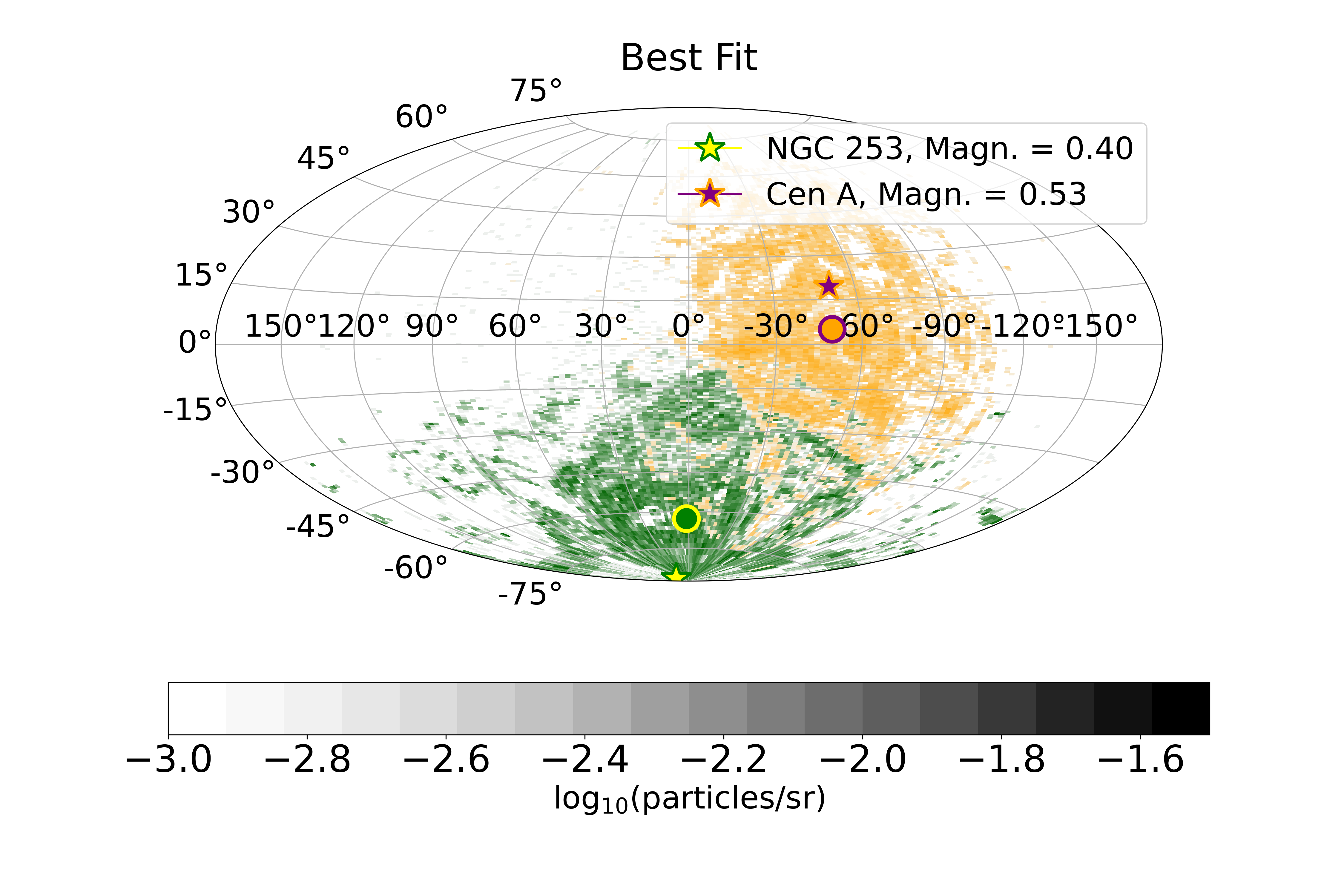}\\
\includegraphics[width=0.49\linewidth]{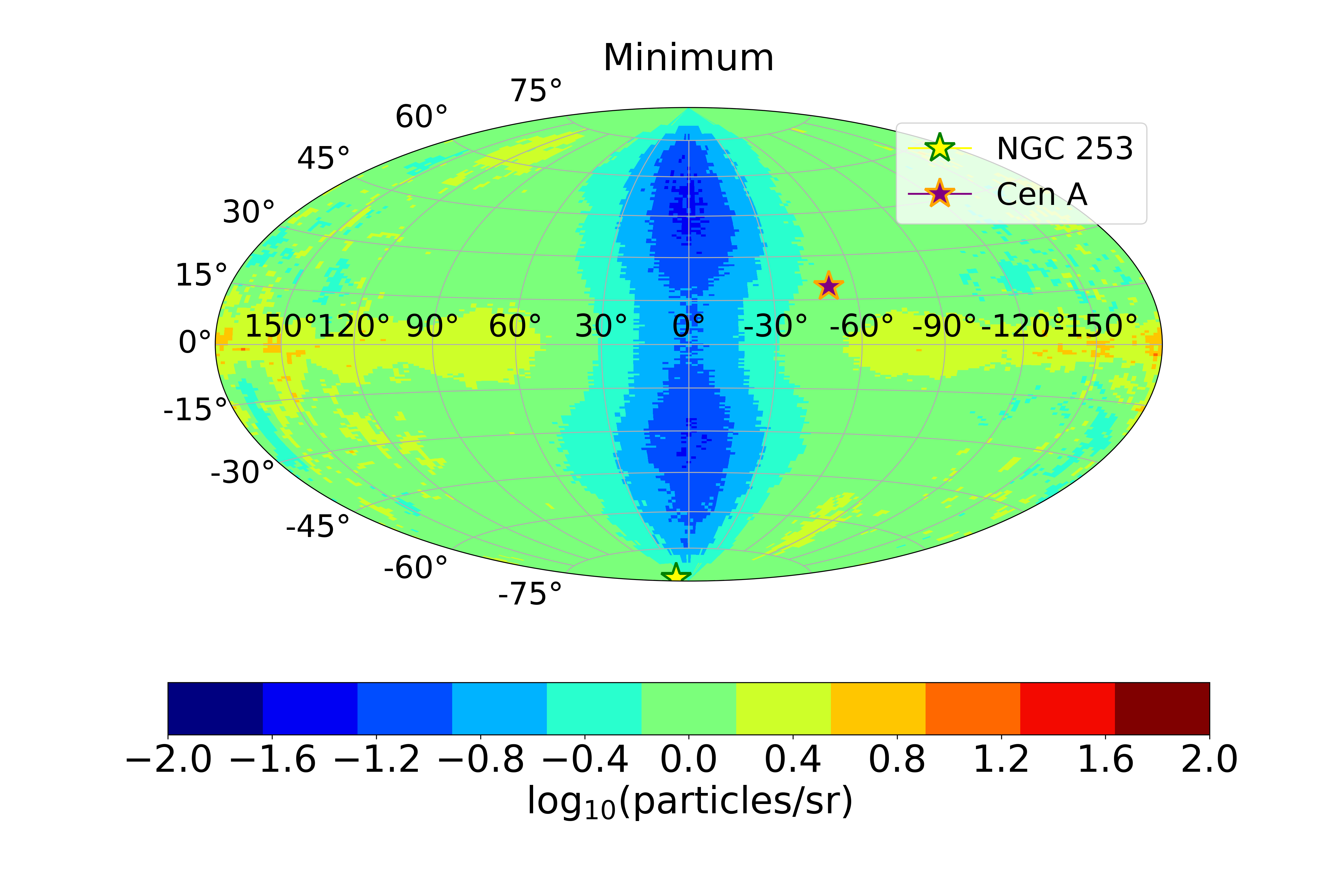}
\includegraphics[width=0.49\linewidth]{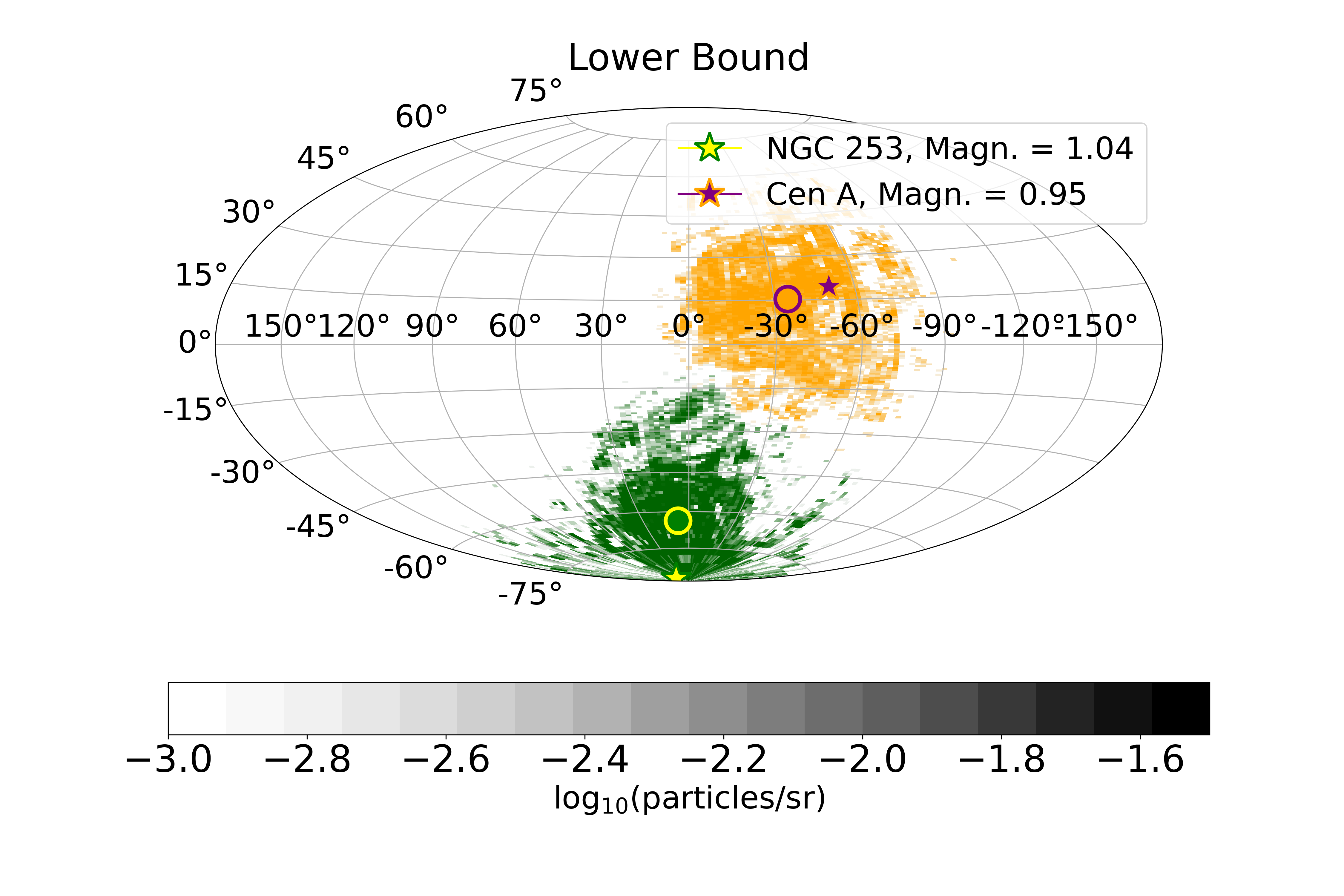}\\
\includegraphics[width=0.49\linewidth]{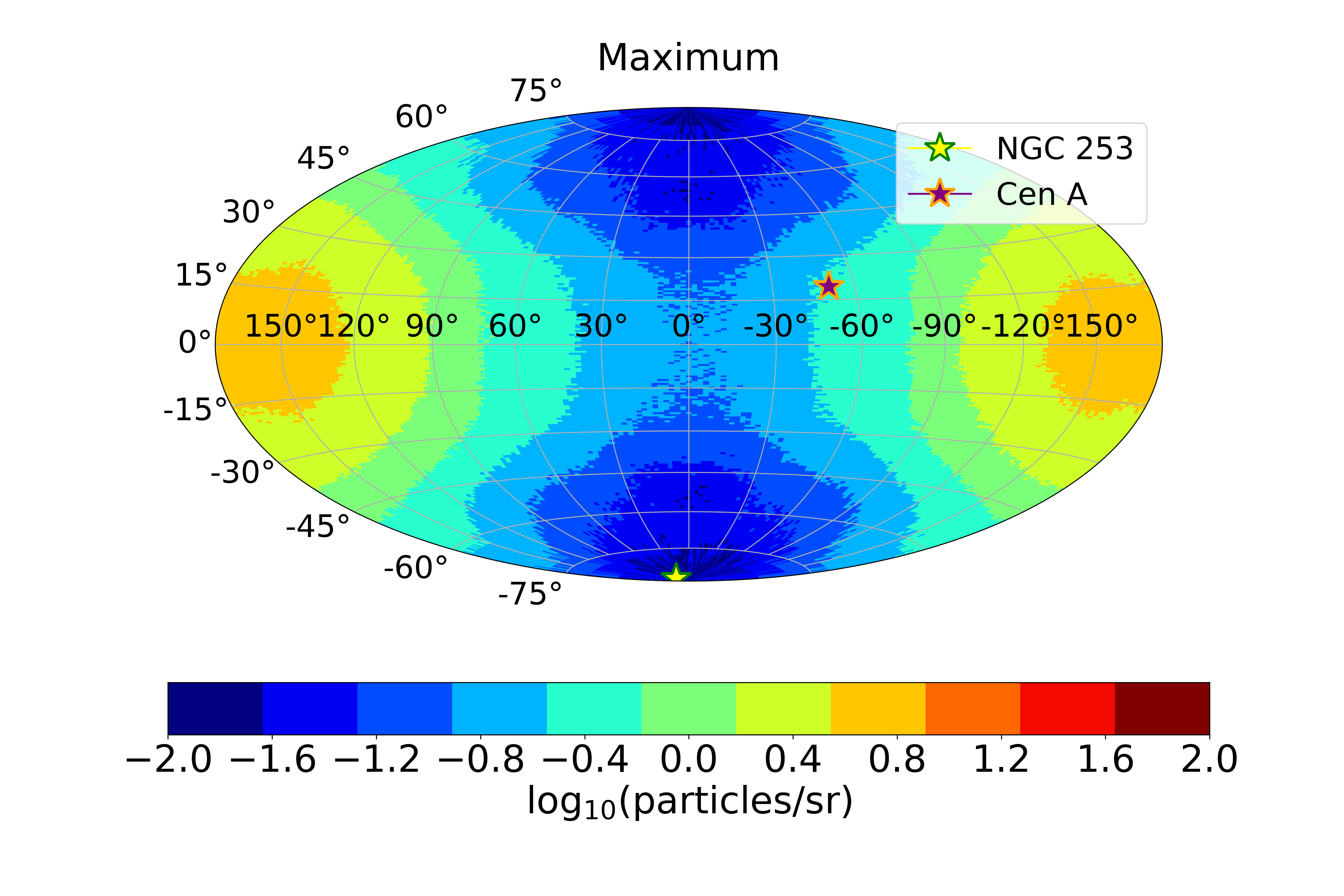}
\includegraphics[width=0.49\linewidth]{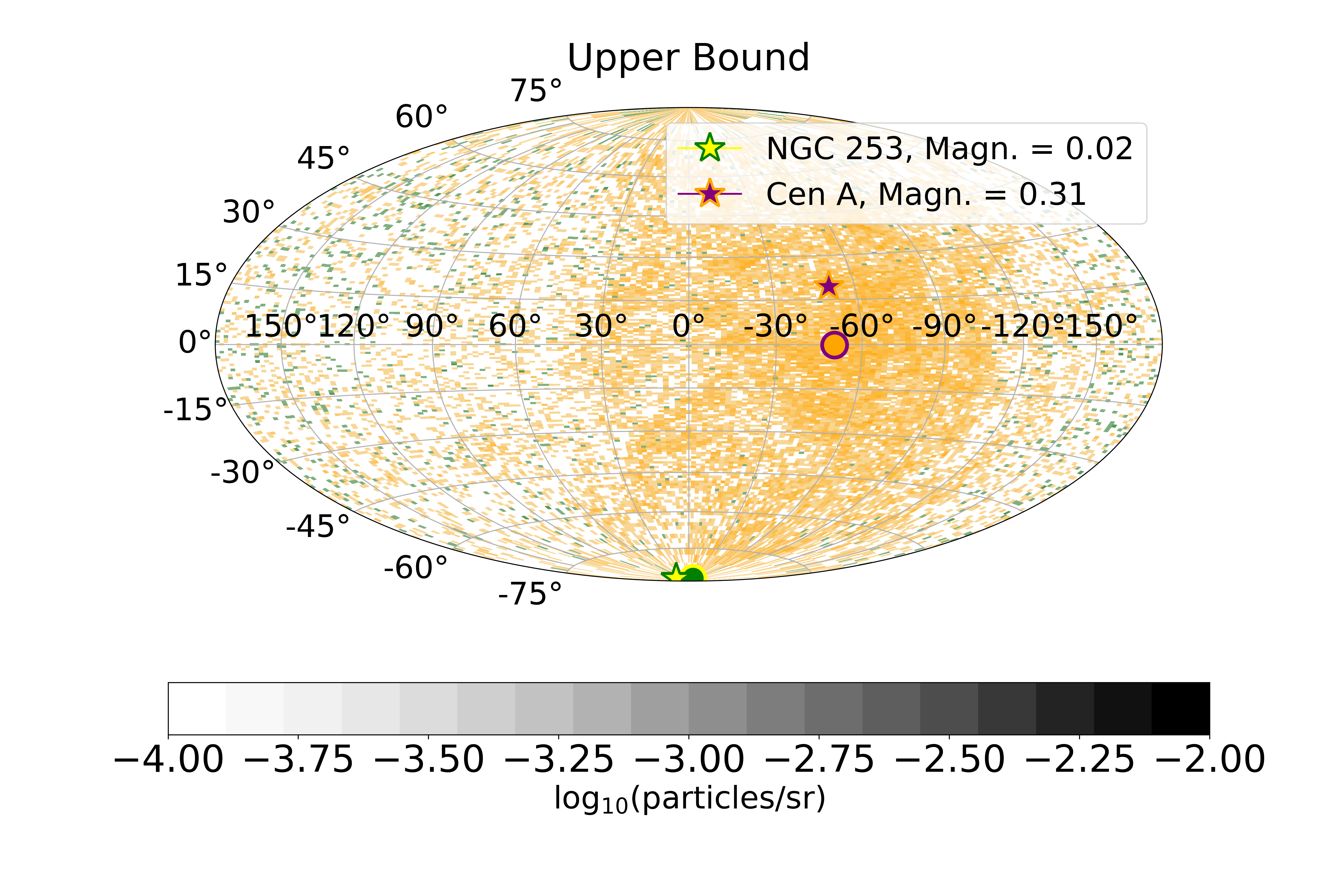}

\hspace*{+9cm}                                      
\caption{{\bf Left:} magnification maps of the extragalactic sky obtained by backtracking an isotropic distribution of cosmic rays (with $R \approx 6 \times 10^{18}$~V) from Earth. These maps are normalised relative to results obtained without magnetic fields with the same total number of events. {\bf Right:} the binned arrival directions of the cosmic rays (with $R \approx 6 \times 10^{18}$~V) from two candidate sources: Cen~A and NGC~253. In the legend we denote for both sources, the ratio ('Magn.') of the number of backtracked cosmic rays within $5^{\circ}$ from the source location for the magnetic field model configuration, to the equivalent number obtained in the absence of magnetic fields. The number of particles in each bin are again normalised by the peak value of a binned histogram obtained without magnetic fields, as represented by the grey colour bars. The mean direction in each plot is denoted by a \tikzcircle[black,fill = gray]{2pt}. \textbf{Top row:} results obtained for best-fit magnetic field parameters, {\textbf{middle row:} lower extreme magnetic field parameters} \& {\textbf{bottom row:} upper extreme magnetic field parameters ($R_{\rm Mag} = $ 7~kpc adopted).}
}

\label{fig:AD_Plots}
\end{figure*}

In figure~\ref{fig:AD_Plots} we show:

\begin{enumerate}
    \item {\bf Left column: } the magnification maps in log(particles/sr), obtained by backtracking an isotropic distribution of cosmic rays from Earth. These magnification maps are made for the best-fit values (\textbf{top}), a set of minimum values (\textbf{middle}) and a set of maximum values (\textbf{bottom}) for the magnetic field model parameter values. 
    To create these histograms we binned the cosmic ray distribution into angular bins (with respect to the Earth) on the escape surface, with 180 bins for both latitudes and longitudes. The histogram values in the maps are normalised to the histogram values obtained for simulations without any magnetic fields present (giving rise to uniform sky brightness). In each map the blue regions denote the areas of the sky where cosmic rays are suppressed and the red regions are the ones where the cosmic rays are enhanced. These magnification maps were made with only our toy model magnetic fields for the Galactic halo bubble region. Note that these deflections of UHECRs can additionally get affected by magnetic fields present at lower Galactic latitudes ($ \rm{|b|}< 15^{\circ}$, in and around the Galactic disc).
    
    \item {\bf Right column: } skymaps for arrival directions
    of cosmic rays from UHECR candidate sources Cen~A and NGC~253. Similar to the above case of the magnification maps we backtrack cosmic rays starting from Earth until they reached an escape radius of 30~kpc from the Galactic centre. The cosmic ray arrival directions from a region of $5^{\circ}$ from the source are then binned.  Like the magnification maps, we normalise these maps to the peak value of the histogram densities in the source region for the case of no magnetic fields.  This gives a normalised value of the number of hits ('Magn.') obtained. Analogous to the magnification maps the \textbf{top}, \textbf{middle} and \textbf{bottom} plots denote the 'Best-fit', 'Minimum' and 'Maximum' cases, respectively. 
    \end{enumerate}

The deflections of UHECRs in the Galactic magnetic field are sensitive to both the structured and turbulent field components of the field in different ways. One of the first effects worth noting is the {\bf suppression effect} for cosmic rays from certain regions of the sky. For the 'Maximum' case, the toy magnetic field model gives rise to the largest suppression factor level in the magnification map, in comparison to the 'Best-fit' and 'Minimum' cases (see Table~\ref{Para_table}). These suppressions and enhancements of UHECRs from different regions of the sky are also seen in the skymaps provided for two potential UHECR sources, namely Cen~A (lon = $\rm -50.49^{\circ}$, lat = $\rm 19.42^{\circ}$) and NGC~253 (lon = $\rm 97.36^{\circ}$, lat = $\rm -87.96^{\circ}$). Because of their positions in the sky, both of these sources lie in the suppression region of the magnification maps for both the 'Best-fit' and 'Maximum' cases. In particular, we note that for NGC~253, the 'Maximum' toy model case leads to a suppression level of $\approx 2\%$ of the level that would arrive for the no magnetic field case, and $\approx 31\%$ for Cen~A.

A second effect introduced by the turbulent magnetic fields is the {\bf spreading effect} (mean spread) of cosmic rays around their originating source direction. In order to quantify this effect, a list is provided below of the mean spread, ($\sigma_{\rm source}$), between the mean direction and the arrival directions of the cosmic rays for the two candidate sources considered:
\newline
\begin{itemize}
        
        \item Best fit - $\sigma_{\rm NGC~253} = {32}^{\circ}$ , {$\sigma_{\rm Cen~A} = {33}^{\circ}$}
        \item Minimum - $\sigma_{\rm NGC~253} = 16^{\circ}$ , $\sigma_{\rm Cen~A} = 21^{\circ}$
        \item Maximum - $\sigma_{\rm NGC~253} = 80^{\circ}$, $\sigma_{\rm Cen~A} = 59^{\circ}$
\end{itemize}

In comparison to the magnitude of these spreading angles, the mean spread of the same rigidity particles from Cen~A and NGC~253 from the JF12 torroidal halo field are  $\sigma_{\rm NGC~253} = 10^{\circ}$ , and {$\sigma_{\rm Cen~A} = 20^{\circ}$}.  The inclusion of the JF12 disc field model to our toy model has a less than $1^\circ$ effect on these results, which is consistent with previous findings \citep{Taylor_2019,Mollerach_2022}.
 
The spread of the cosmic rays obtained for our toy model are therefore potentially considerably larger (up to 5-10 times bigger) than those obtained for the JF12 toroidal halo. The primary driver of this difference is that our toy model possesses a larger level of turbulent magnetic fields than structured fields. It is also worth noting that the mean spread of Cen A ($\sigma_{\rm Cen~A} = {33}^{\circ}$) for the 'Best-fit' case of our toy model is comparable with the PAO observations~{\citep{Auger_ICRC_2021,Auger_2022}}. 

Additional to this spreading effect, the presence of a structured field component in the magnetic field model leads to the {\bf coherent deflection} of the mean direction of the ensemble of cosmic rays away from the source direction. Following the propagation of cosmic rays from the two candidate sources NGC~253 and Cen~A, the mean shifted positions (lon, lat) for the three cases are as follows:
\begin{itemize}
    \item Best fit - NGC~253: ($2^{\circ}$,$-63^{\circ}$) \& Cen~A: ($-50^{\circ}$,$5^{\circ}$) 
    \item Minimum - NGC~253: ($7^{\circ}$,$-63^{\circ}$) \& Cen~A: ($-35^{\circ}$,$15^{\circ}$) 
    \item Maximum - NGC~253: ($-39^{\circ}$,$-88^{\circ}$) \& Cen~A: ($-50^{\circ}$,$-0.2^{\circ}$) 
\end{itemize}

In comparison, the mean shifted positions from the JF12 toroidal halo model for the two sources are  NGC~253: ($-0.8^{\circ}$,$-34^{\circ}$) and Cen~A: ($-44^{\circ}$,$-1^{\circ}$). The mean shifted positions from the XH19 model for NGC~253 are: ($1.2^{\circ}$,$-29^{\circ}$) and Cen~A: ($-48^{\circ}$,$-5.1^{\circ}$).

Additionally, for the case of cosmic rays from NGC~253, an interesting difference between our toy model and the JF12 toroidal halo model is worth noting. For the JF12 toroidal halo model (see Appendix~\ref{Appendix_D}), the mean shifted position of cosmic rays from NGC~253 is situated at roughly a latitude of $-34^{\circ}$ (also seen in \citet{Arjen_2021}). In contrast to this, in our 'Best-fit' toy model case this value is at approximately $-63^{\circ}$, which would be in better agreement with the PAO observations~\citep{Auger_Starburst2018,Auger_2022} if this southern Galactic hemisphere hotspot does indeed originate from NGC~253. 
The presence of stronger turbulent magnetic fields in comparison to the structured fields in our toy model results in a relatively small coherent deflection and large spreading of cosmic rays from NGC~253. The contrary is applicable to the JF12 toroidal halo model, which has stronger structured fields and a weak turbulent magnetic field model that results in the large latitude shift of cosmic rays from NGC~253.

The suppression, spreading and coherent deflection effects place challenges on the association of cosmic rays to their originating sources. It can be seen that the 'Maximum' toy model magnetic field would make associating cosmic rays to their source extremely challenging at the energies considered, whereas the 'Best-fit' or 'Minimum' cases make this possible. This is due to the fact that the structured fields are responsible for the overall direction of the particle deflections, whereas turbulent fields are responsible for spreading out the directions of the particle deflections around this overall deflected direction. For cases in which the turbulent magnetic field component dominates, and this field strength component is large, the source directions can be completely washed-out. This washing-out of the source association is evident for the 'Maximum' case in Fig.~\ref{fig:AD_Plots}, cosmic rays from sources like NGC~253 are largely deflected from the source position by the magnetic field structure with a mean spread of $\sigma_{\rm NGC~253} = 80^{\circ}$. Likewise, in the case of Centaurus~A the final positions are spread out over a large region of the sky with a mean spread of $\sigma_{\rm Cen~A} = 59^{\circ}$, making association with the source position challenging. From both the magnification and arrival direction maps in Fig.~\ref{fig:AD_Plots}, it is evident that the best-fit and lower extreme ('Minimum') parameters allow some degree of association of the deflected UHECRs with their original source position. However, in the upper extreme ('Maximum') parameters, such a connection between the point of origin of cosmic rays and their final positions is heavily erased. 

Apart from for the nitrogen nuclei, we also computed the deflections of ultra high energy cosmic ray protons at 40~EeV (see Appendix~\ref{Appendix_E} for skymaps). 
With mean spreads for the best-fit model of $\sigma_{\rm NGC~253} = 5^{\circ}$ and $\sigma_{\rm Cen~A} = 8^{\circ}$, the deflections for the proton scenario are significantly reduced compared with the nitrogen case. Such skymaps for the arrival directions of different nuclear species groups can therefore be helpful in studying the effect of the magnetic fields in the Galactic halo bubbles on different cosmic ray species. Studies like these can be useful for future ultra high energy cosmic ray observatories like AugerPrime \citep{Auger_Prime_2016, Auger_Prime_2019}, an upgrade to Auger, which will be able to identify UHECR composition on an event by event basis.

\section{Conclusions \& Outlook}
\label{Conclusions}
The radio observations \citep{WMAP_2004,Carretti_2013,Planck_Haze_2013} strongly indicate the presence of non-thermal particles at high Galactic latitudes. We consider a toy model for the Galactic halo bubble magnetic field, in an effort to best describe the polarised synchrotron emission seen in the Planck 30~GHz data from the Galactic halo bubble region.

Utilising our toy model for the Galactic halo {bubbles} magnetic field, and making comparisons of the polarised synchrotron emission predicted by it to the Planck 30~GHz data, we explore the region of model parameters capable of providing a good description of the data. Significant evidence is found for the presence of an extended magnetic field component out in the Galactic halo bubble region. 
For the best-fit case, we obtain a large value of $\sim 7~\mu$G for the total magnetic field (dominated by the turbulent magnetic field $B_{\rm tur} = 6.72~\mu$G) in the Galactic
halo bubble region with a large spatial extent to $\sim 6$~kpc in height above the Galactic disc. These results are consistent with the magnetic field strength estimates \citep{Carretti_2013} and the spatial extension scales \citep{Su_2012}.

The total magnetic field content in the halo bubble region from our best fit parameter set is $\approx 10^{55}$~ergs. We note that this value is somewhat smaller than the observational inference made from the eROSITA measurements~\citep{eROSITA}, which suggested the presence of some $10^{56}$~ergs of thermal particles from somewhat larger scale Galactic halo bubbles.
In comparison to these energy contents, the total magnetic field energy content in the halo field component of the JF12 model is $4\times 10^{54}$~ergs and $3\times  10^{54}$~ergs for the toroidal halo and X-field respectively \citep{Taylor_2019}.

Using the maximum and minimum constraints on the magnetic field model parameter values we obtained, the range of deflection that UHECRs experience in passing through such Galactic halo bubble magnetic field structure was subsequently investigated. A significant range in predictions of both: a) the magnification of different regions of the extragalactic sky, and b) the deflection of cosmic rays arriving from different local extragalactic sources was found. 
For the best-fit case from our magnetic field model we obtain mean spreads, $\sigma$, of $\sim 30^{\circ}$ for both NGC~253 and Cen~A, this has been found to be 
consistent with PAO observations~\citep{PierreAuger_2014, Auger_2022}, for both the hotspots around Cen~A and NGC~253.

Beyond synchrotron emission inferences of the halo bubble magnetic field, Faraday rotation measure observations also promise to offer new insight. Despite the  growing body of indirect evidence pointing to the existence of diffuse hot thermal gas at high Galactic latitudes \citep{Gupta_2012,Hodges_Kluck_2016}, its presence has been difficult to probe. This gas, however, has  recently been detected by the discovery of the eRosita bubbles \citep{eROSITA}. Faraday rotation from these high latitude thermal electrons can also be utilised in the future to probe magnetic fields in the Galactic halo bubbles.

Of particular promise are future observations of both the dispersion measure and Faraday rotation measure from fast radio bursts (FRBs). Collectively, these can also provide information about both the thermal gas density and the magnetic fields out in the halo bubbles, in an analogous manner to the use of pulsars to probe the Galactic disc magnetic field. Although presently the distribution of FRBs in the Galactic halo bubble region remains too sparse to allow such a study, this situation is expected to improve with the upcoming data releases from FAST and MeerKat \citep{FRBs_dis_2021}.

Additionally, further insights into the halo magnetic field structure are also anticipated from the use of ultra high energy cosmic rays from potential local sources \citep{Lemoine_2009,Liu_2013,Arjen_2021}.
With the anticipated completion of AugerPrime, the cosmic ray composition on an event-by-event basis promises to probe the deflection of UHECR in a rigidity dependent manner \citep{Auger_Prime_2016,Auger_Prime_2019}, allowing present correlations to candidate objects to be further tested.

\section*{Data availability}
This study was done using the publicly available Planck data \hyperlink{Planck}{http://pla.esac.esa.int/pla/}. The codes used for the cosmic ray propagation was CRPropa 3 \citep{CRPropa3_2016}\footnote{\textcolor{purple}{https://crpropa.github.io/CRPropa3/index.html}} which is also a publicly available software. The codes used for the generation of synthetic synchrotron maps can be made available upon request to the corresponding author. 

\section*{Acknowledgements}
The authors acknowledge support from DESY (Zeuthen, Germany), a member of the Helmholtz Association HGF. V.~Shaw would like to thank Mike Peel from the Planck Collaboration for helpful discussions about the Planck data. 

\bibliographystyle{mnras}
\bibliography{references.bib}

\appendix

\section{Turbulent field wavelength range}

\label{Appendix_A}
We adopt a narrow wavelength range for the turbulent fields, $L_{\rm min}=200$~pc and $L_{\rm max}=400$~pc. This was mainly done due to computational limitations and is similar to what was adopted by \citet{West_Helicity}. The effect of this truncation on the turbulence power spectra can be quantified. The energy density in the turbulent modes is given by:
\begin{equation}
    \delta B^{2} = \int _{L_{\rm min}}^{L_{\rm max}} \frac{\delta B^{2}}{\rm{d}L}{\rm{d}}L = \frac{B_{0}^{2}}{L_{\rm max}}\int  _{L_{\rm min}}^{L_{\rm max}}\left(\frac{L}{L_{\rm max}}\right)^{q-2} dL
\end{equation}
\begin{equation}
  \delta B^{2}  =\frac{B_{0}^{2}}{q-1}\left[1-\left(\frac{L_{\rm min}}{L_{\rm max}}\right)^{q-1}\right] 
\end{equation}
Usually (for typical values of the turbulence cascade index, $q$,
considered) this integral is insensitive to $L_{\rm min}$ , being dominated by the longest mode values, so where one truncates the lower end of the integral has only a small effect. The artificial enhancement of $B_0^2$ due to the truncation that we adopt is $\left[1-(1/2)^{q-1}\right]^{-1}$, which for $q=5/3$ is 2.7, corresponding to $B_0$ being artificially enhanced by $\sim $60\% due to the early truncation of the turbulent spectrum adopted.

\section{Turbulent field spectrum}

\label{Appendix_B}
\begin{figure*}
    \includegraphics[width = 0.49\linewidth]{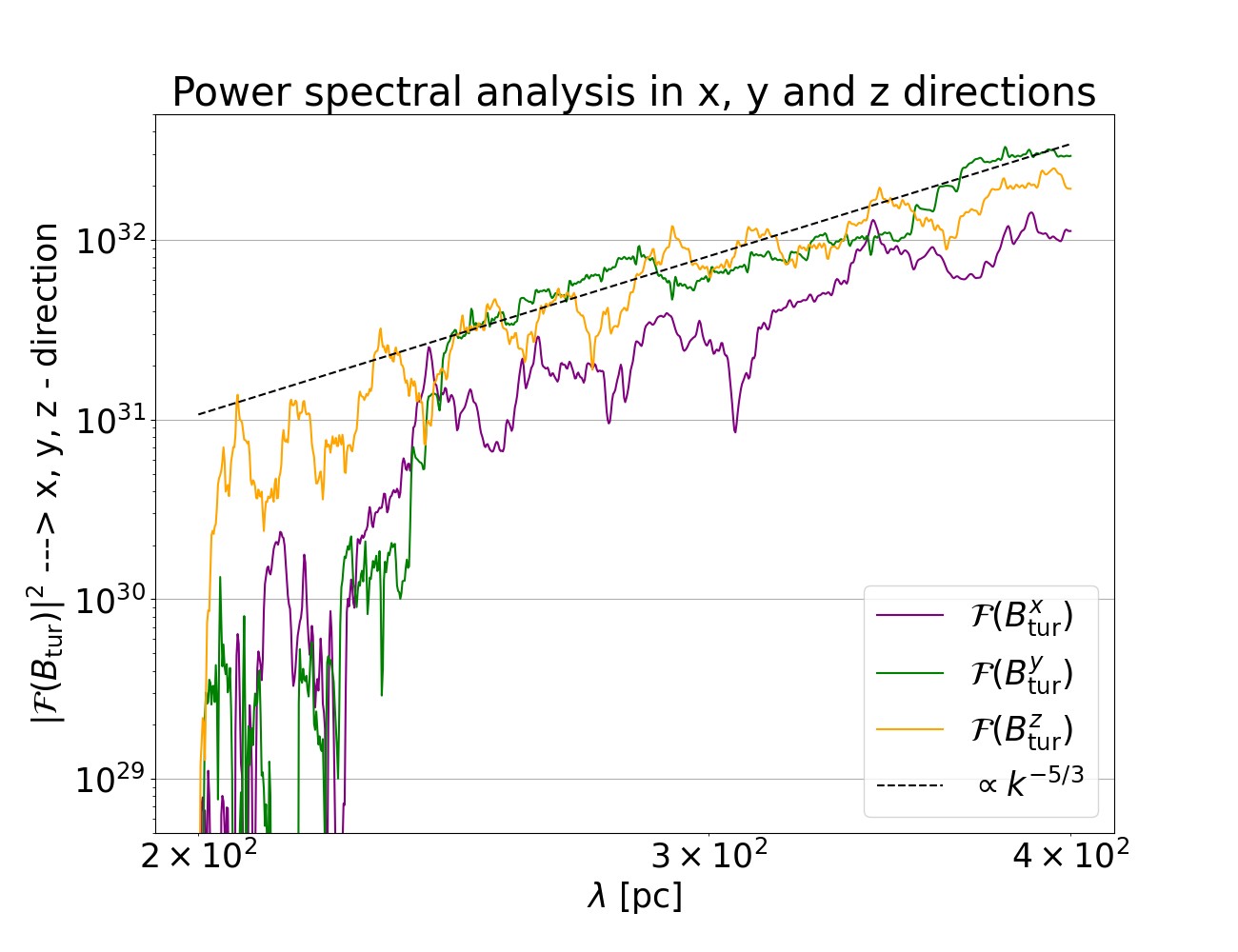}
    \caption{Power spectra of turbulent magnetic fields, evaluated along three orthogonal directions, namely the $x$, $y$ and $z$ directions.}
    \label{fig:PowerSpectrum}
\end{figure*}

As discussed in Section~\ref{GMF}, we generate the turbulent fields for our model using CRPropa~3~\citep{CRPropa3_2016}. The minimum and maximum wavelength we use to generate these fields are 
$L_{\rm min}$ = 200 pc and $L_{\rm max}$ = 400 pc and $L_{\rm coh} \approx $~150 kpc. One of the major reasons why we do not have more decades covered for the wavelength is because of the time it takes to generate these fields using CRPropa. 
We investigated power spectra for different realisations of the turbulent field. In Fig.~\ref{fig:PowerSpectrum}, we plot power spectra in $x$, $y$ and $z$ directions, after averaging over the other two directions. We chose a step size of 1~pc and integrate up to $\approx 9\times10^4$~pc. We chose this particular realisation since it followed closely a power-law spectrum of index 5/3, with a similar amount of power in each direction (i.e.~was reasonably isotropic). 

\section{Polarised Synchrotron Emission}
\label{Appendix_C}
\begin{figure*}
\centering
\includegraphics[width = 0.49\linewidth]{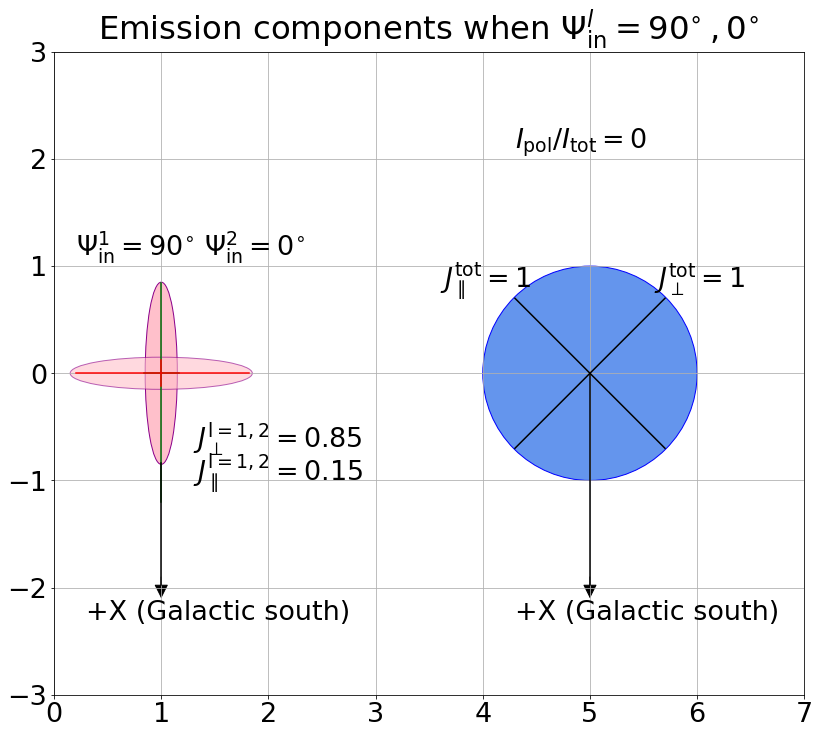}
\label{fig_tot_intensity}
\includegraphics[width = 0.49\linewidth]{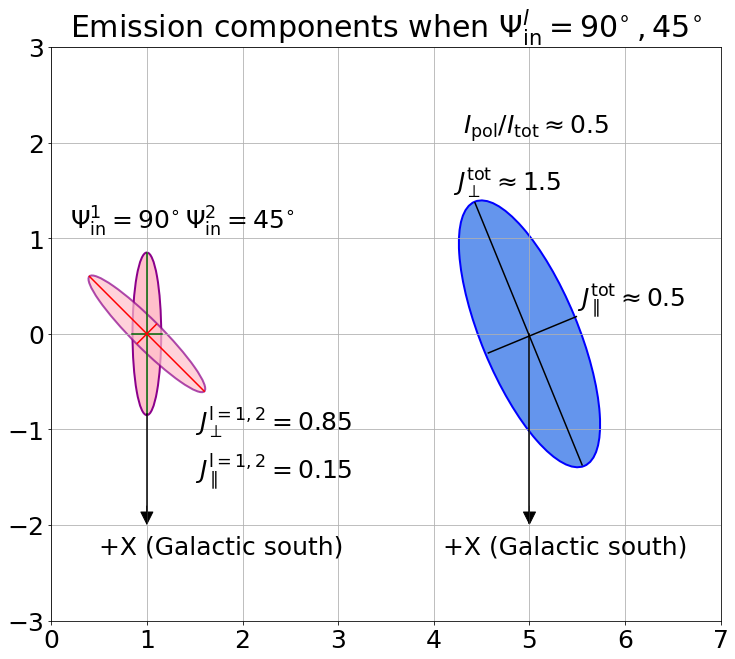}
\label{fig_pol_intensity}
\caption{Diagram depicting visually the resultant ellipse (blue), obtained from the summation of two ellipses of equal magnitude (pink), but different orientations. The resultant ellipse dictates the resultant total and polarised intensities.}
\label{fig_tot_pol_intensity}
\end{figure*}

As discussed in Section~\ref{Synchrotron_theory}, the line-of-sight components for Stokes parameters are given by:
\begin{eqnarray}
Q^l_{\rm in} = ({J_{\perp}^l} - J_{\parallel}^l) \ {\cos}(2\Psi^l_{\rm in}) \ ,\\ U^l_{\rm in} = ({J_{\perp}^l} - J_{\parallel}^l) \ {\sin}(2\Psi^l_{\rm in}) \ .
\end{eqnarray}
We can take two test cases (I \& II) given in the left and right panel of Fig.~\ref{fig_tot_pol_intensity}, respectively. We consider that there are two steps along a line of sight for which ${J_{\perp}^{(1,2)}}$ = 0.85 and $J_{\parallel}^{(1,2)}$  = 0.15. 

In case I the angles $\Psi_{\rm in}^{(1,2)}$ = $90^{\circ}$ \& $0^{\circ}$. The resultant value of $I_{\rm pol}$ = 0 by virtue of Eq.~\ref{eq_I_pol} and we only have a contribution to $I_{\rm tot}$. This implies that for case I the resultant emission is seen only in total intensity, since the values of $J_{\perp}^{\rm tot}$ = $J_{\parallel}^{\rm tot}$. 

In case II we apply similar calculations to case I, however, now the angles are $\Psi_{\rm in}^{(1,2)}$ = $90^{\circ}$ \& $45^{\circ}$. This, in turn, results in contributions to both polarised emission $I_{\rm pol}$ and total intensity $I_{\rm tot}$. Thus, the values of $J_{\perp}^{\rm tot} \neq J_{\parallel}^{\rm tot}$. In the right panel of Fig.~\ref{fig_tot_pol_intensity}, we only show the polarised intensity for simplicity, however, there will be both total intensity and polarised intensity present.

\section{Polarised synchrotron emission from other halo models}
\label{Appendix_F}
\begin{figure*}
\centering
\includegraphics[width=0.49\linewidth]{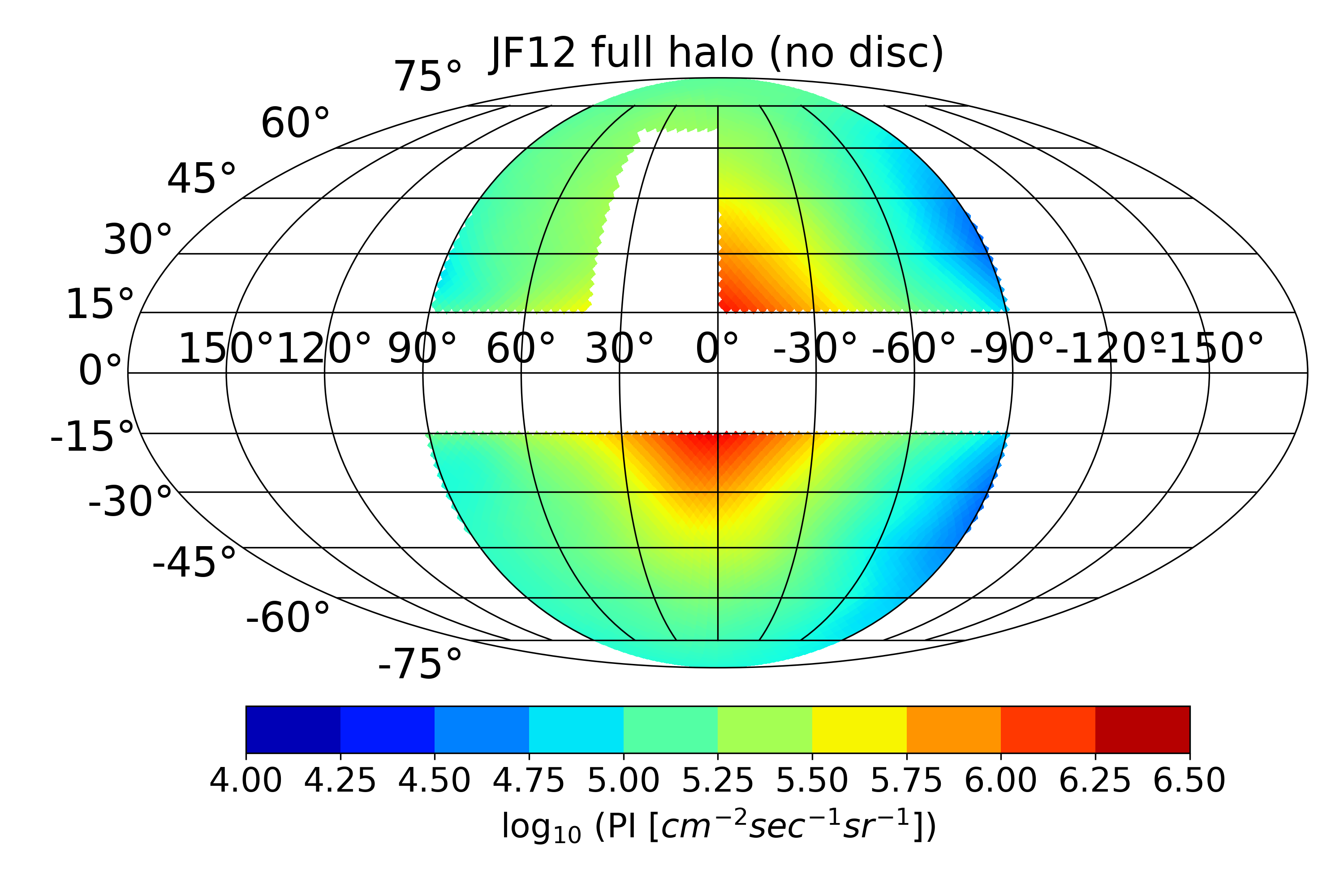}
\includegraphics[width =0.49\linewidth]{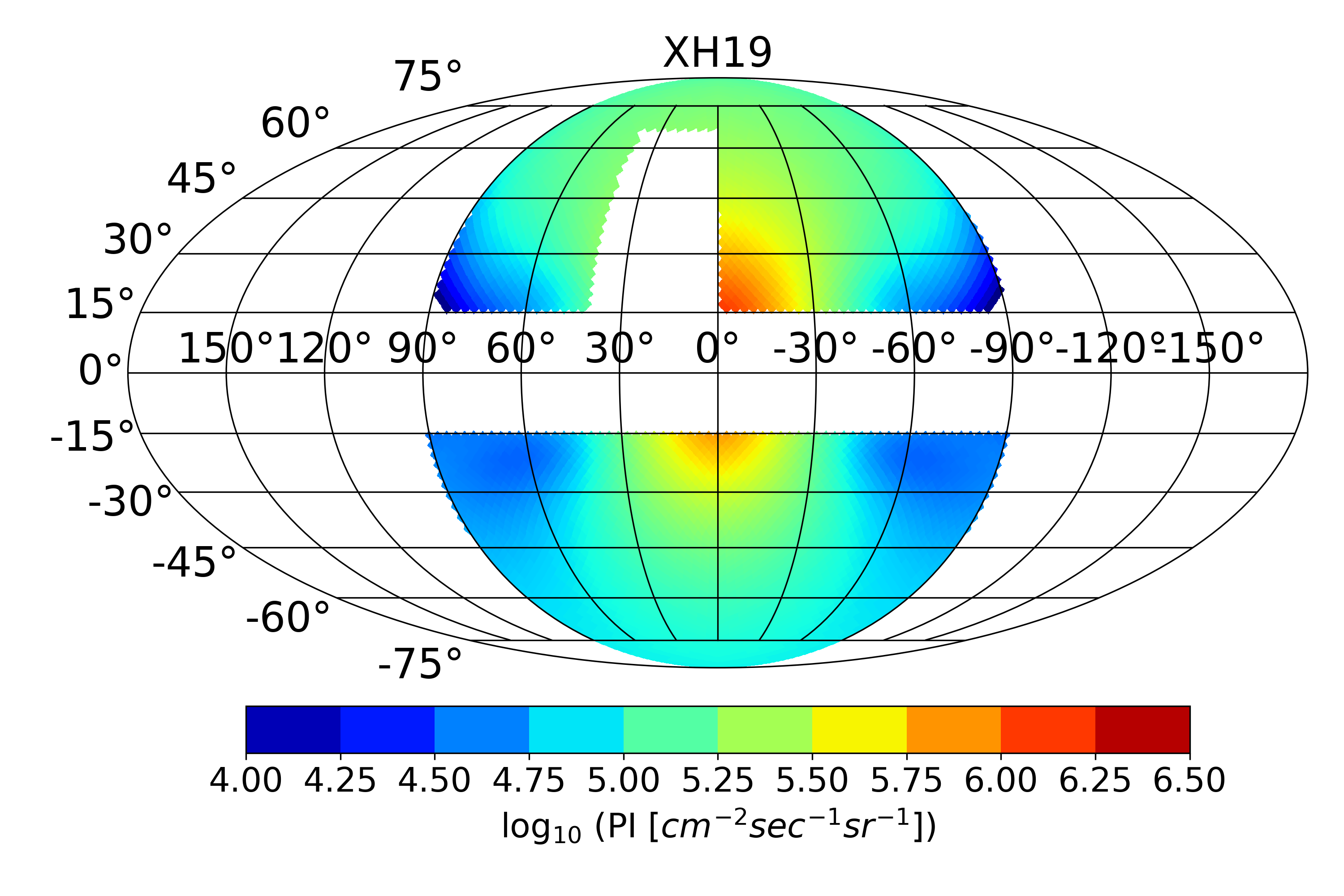}%
\caption{\textbf{Top:} Simulated polarised intensity for the JF12 full halo (no disc) (\textbf{left}) and XH19 model (\textbf{right}) with the same electron distribution (see eq.\ref{Eq_WMAP_EdNdE}), smoothing and angular cuts as in the results in Fig.~\ref{fig:Skymaps}.}
\label{fig:Skymaps_XH19_JF12}
\end{figure*}

Polarised synchrotron emission from our toy model gives ${\chi^2}/{\rm d.o.f} = $~1.7 when compared with Planck polarised 30~GHz data (see Fig. \ref{fig:Skymaps}) for the best-fit parameter values. In figure \ref{fig:Skymaps_XH19_JF12}, we show polarised synchrotron emission from the XH19 and JF12 full halo (no disc) models with the same electron distribution (see eq. \ref{Eq_WMAP_EdNdE}), smoothing and angular cuts as we used for the polarised synchrotron emission from our toy model. We compare these models with the Planck polarised 30~GHz data for the same region and find 
a  ${\chi^2}/{\rm d.o.f} = $~11.0 for the XH19 model and  ${\chi^2}/{\rm d.o.f} = $~6.0 for the JF12 full halo model. 
Both these models fit the data poorly due to either their weak turbulent magnetic field model as seen in the case of JF12 full halo or the complete lack of any turbulent magnetic fields like in the XH19 model. 
We conclude from these comparisons that our model is statistically better able to describe the high latitude polarised synchrotron emission, seen in the Planck 30 GHz polarised data, than the JF12 full halo and the XH19 model.

\section{Arrival directions from other halo models for nitrogen at E = 40~E\lowercase{e}V}
\label{Appendix_D}
\begin{figure*}
\centering
\includegraphics[width=0.60\linewidth]{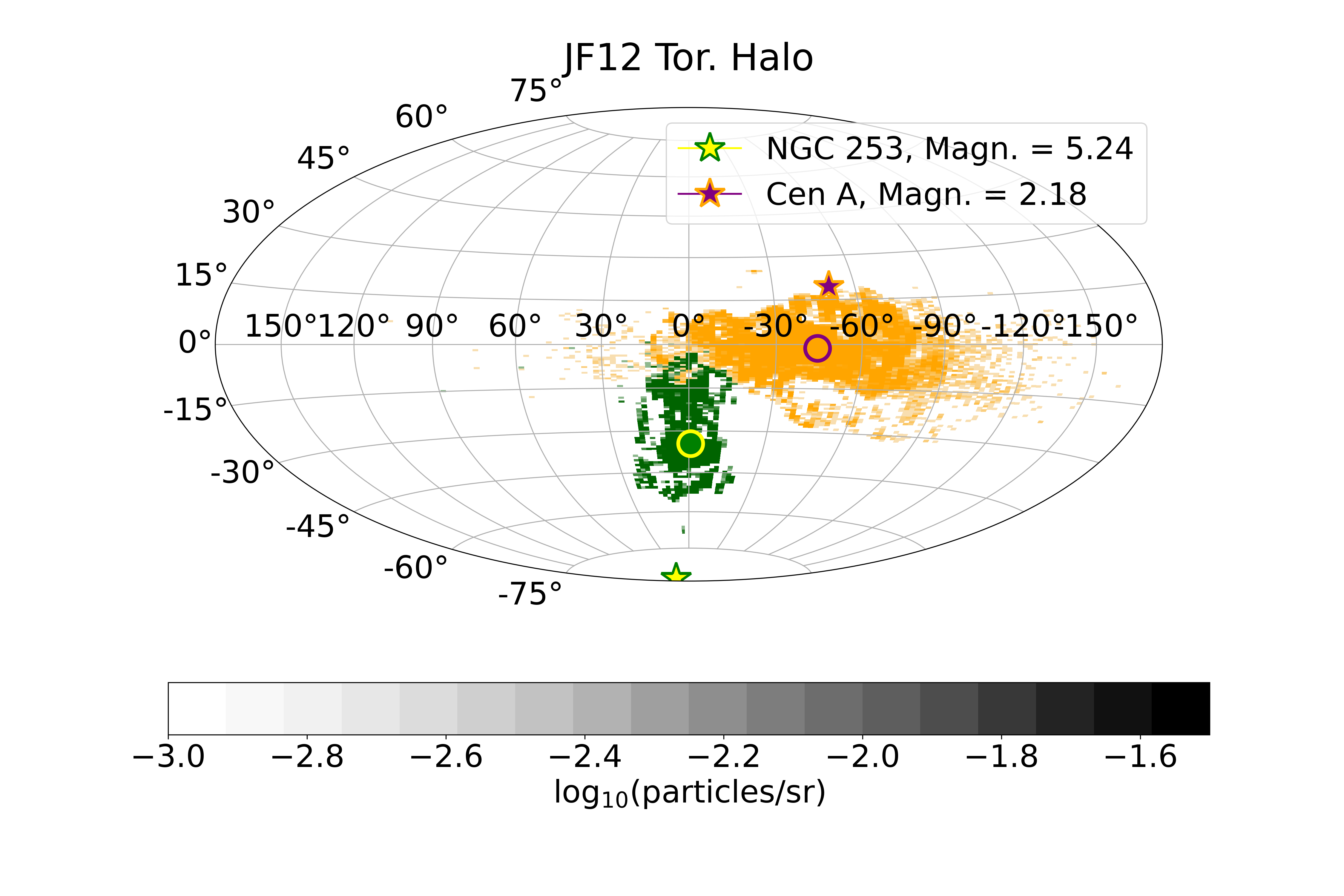}
 \caption{Arrival direction map of cosmic rays (with $E \approx 40 \times 10^{18}$~eV) deflected from the JF12 toroidal halo model for two potential UHECR sources. It can be seen that the mean direction of the deflection for the JF12 toroidal halo is at a higher latitude than for the toy model, shown in Fig.~\ref{fig:AD_Plots}.}
\label{JF12_AD}
\end{figure*}

We calculate the arrival directions of cosmic rays (nitrogen at 40~EeV) for two candidate sources, Cen~A and NGC~253, for the JF12 toroidal halo model, shown in Fig.~\ref{JF12_AD}. We normalise these binned arrival directions by the peak value of the histogram obtained from the same setup without magnetic fields present. 
The mean shifted positions obtained from the JF12 toroidal halo model for the two sources are ($-0.8^{\circ}$,$-34^{\circ}$) for NGC~253 and ($-44^{\circ}$,$-1^{\circ}$) for Cen~A, and the mean spread from Cen~A and NGC~253 are  $\sigma_{\rm NGC~253} = 10^{\circ}$ , and {$\sigma_{\rm Cen~A} = 20^{\circ}$}.
It can be seen from Fig.~\ref{JF12_AD} that the JF12 toroidal halo displaces the binned arrival directions to much higher latitudes in the case of NGC~253. This is because the structured field strength in the JF12 toroidal halo is stronger than the turbulent field and hence the mean deflection from the source position is larger.

We applied the same method as well to study the UHECR deflections (nitrogen at 40~EeV) for the XH19 model. In this case, the mean shifted positions obtained for NGC~253 and Cen~A are ($1.2^{\circ}$,$-29^{\circ}$) and ($-48^{\circ}$,$-5.1^{\circ}$) , respectively.
The high latitude deflections obtained from the XH19 model is due to the presence of structured magnetic fields which results in only the coherent deflection cosmic rays. The absence of turbulent fields results in a negligible mean spread in the UHECR arrival directions.
\section{{Arrival directions of UHECR protons at E = 40~E\lowercase{e}V}}
\label{Appendix_E}
{We calculate the arrival directions of cosmic rays for two candidate sources, Cen~A and NGC~253, for protons at 40~EeV for the best-fit case and the upper-bound (maximum) case (see Fig.~\ref{fig:proton_skymaps}) of our toy model as magnetic field model. We obtain the following mean spreads:
\begin{itemize}
        \item Best fit - $\sigma_{\rm NGC~253} = 5^{\circ}$ , {$\sigma_{\rm Cen~A} = 8^{\circ}$},
        \item Maximum - $\sigma_{\rm NGC~253} = 12^{\circ}$, $\sigma_{\rm Cen~A} = 17^{\circ}$.
\end{itemize}
The mean shifted positions after the coherent deflection through the toy model magnetic fields are:
\begin{itemize}
    \item Best fit - NGC~253: ($10^{\circ}$,$-83^{\circ}$) \& Cen~A: ($-47^{\circ}$,$20^{\circ}$), 
    \item Maximum - NGC~253: ($6^{\circ}$,$-61^{\circ}$) \& Cen~A: ($-30^{\circ}$,$17^{\circ}$). 
\end{itemize}

\begin{figure*}
\centering
\includegraphics[width=0.49\linewidth]{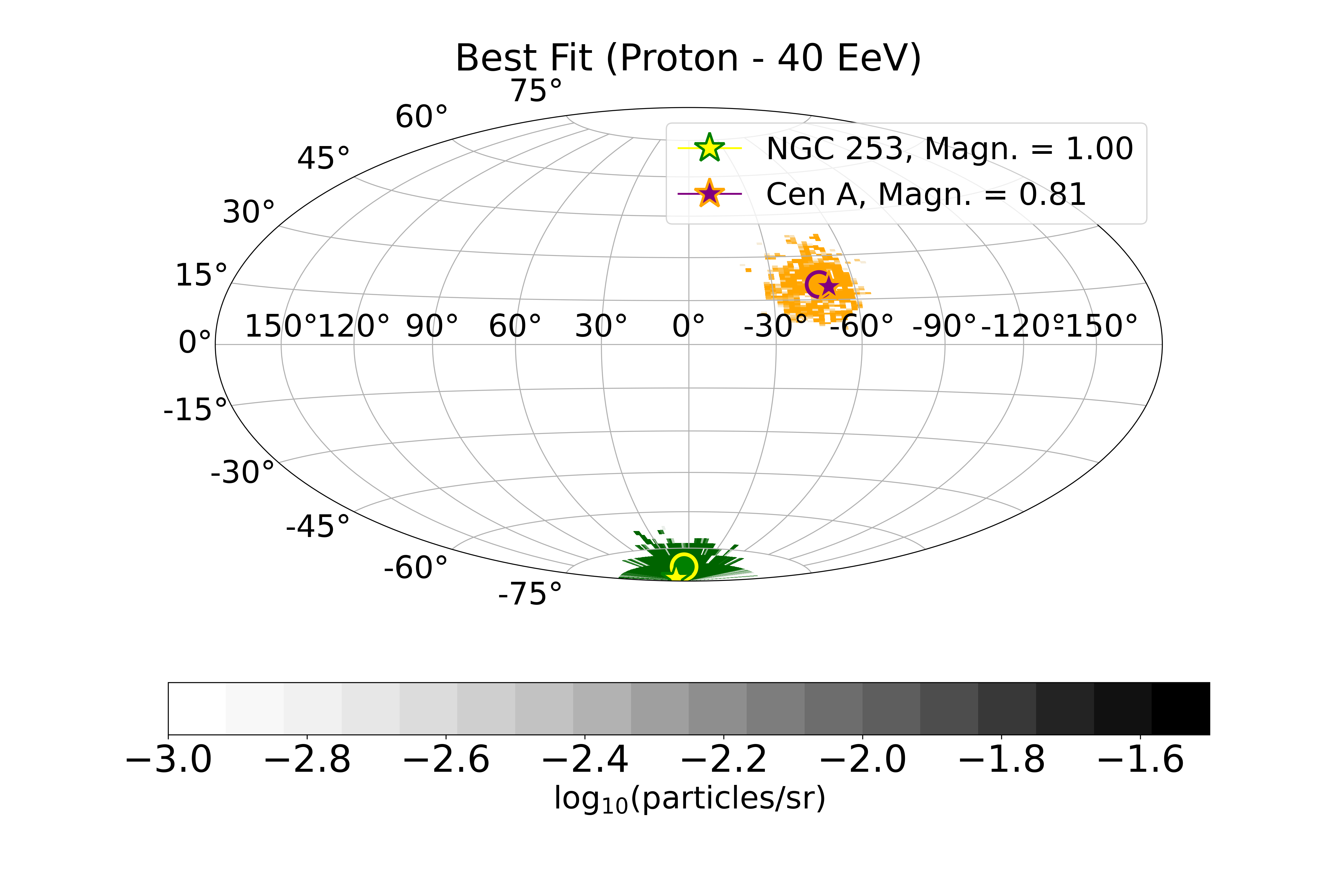}
\includegraphics[width=0.49\linewidth]{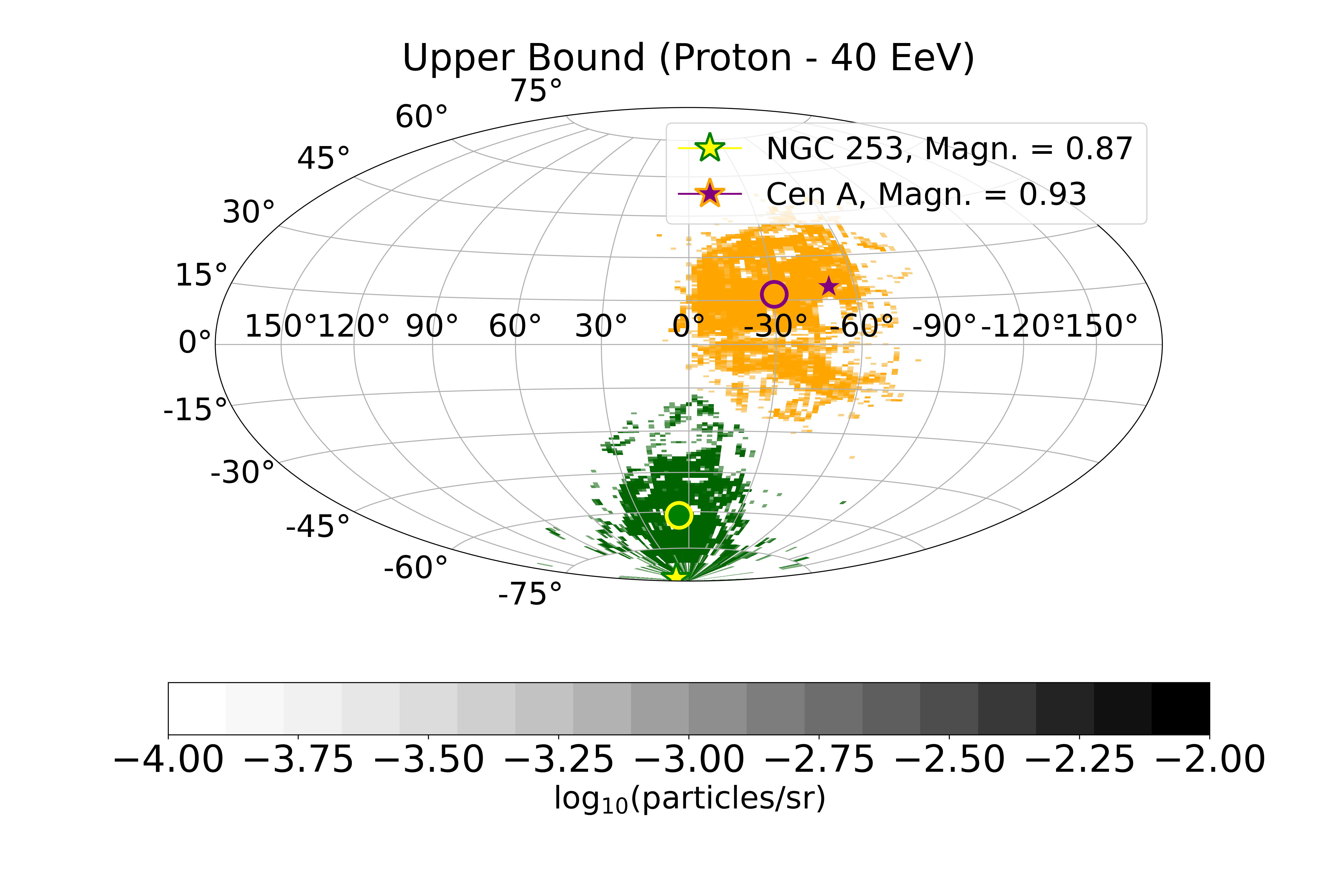}
 \caption{Arrival direction map of cosmic ray protons (with $E \approx 40 \times 10^{18}$~eV) for the best fit case (\textbf{left}) and upper bound case (\textbf{right}) of our toy model. }
\label{fig:proton_skymaps}
\end{figure*}
}

\end{document}